\documentclass[aps,prb,twocolumn,epsf,epsfig,amsmath]{revtex4}
\usepackage[toc,page]{appendix}
\usepackage{hyperref}
\usepackage{graphicx,pdflscape,placeins}% Include figure files
\usepackage{color}
\usepackage{bm}% bold math
\usepackage{alltt,dsfont}
\usepackage{appendix}
\usepackage{amsmath,amssymb}
\usepackage{braket}
\usepackage[capitalise]{cleveref}
\usepackage{ulem}

\usepackage{atveryend}
\makeatletter
\let\origcitation\citation
\AtEndDocument{\def\mycites{}%
  \def\citation#1{\g@addto@macro\mycites{#1^^J}\origcitation{#1}}}
\AtVeryEndDocument{\newwrite\citeout\immediate\openout\citeout=\jobname.cit
  \immediate\write\citeout{\mycites}\immediate\closeout\citeout}
\makeatother

\newcommand {\apgt} {\ {\raise-.5ex\hbox{$\buildrel>\over\sim$}}\ }
\newcommand {\aplt} {\ {\raise-.5ex\hbox{$\buildrel<\over\sim$}}\ }
\newcommand{\lessim}{\aplt}
\newcommand{\gssim}{\apgt}

\newcommand{\tx}[1]{\text{#1}}
\newcommand{\iden}{ \mathds{ 1}}
\newcommand{\G}{{\cal{G}}}
\newcommand{\GH}{{\bf g}}
\newcommand{\GHI}{\GH^{-1}}

\newcommand{\V}{{\mathcal V}}

\newcommand{\si}{\sigma}

\newcommand{\tJ}{$t$-$t'$-$J$ \ }
\newcommand{\beq}{\begin{equation}}
\newcommand{\eeq}{\end{equation}}
\newcommand{\barray}{\begin{eqnarray}}
\newcommand{\earray}{\end{eqnarray}}
\def\bal#1\eal{\begin{align}#1\end{align}}
\newcommand{\nn}{\nonumber}

\newcommand{\disp}[1]{Eq.~(\ref{#1})}
\newcommand{\dispop}[1]{~{\bf (\ref{#1})}}
\newcommand{\refdisp}[1]{Ref.~[\onlinecite{#1}]}
\newcommand{\figdisp}[1]{Fig.~\ref{#1}}

\newcommand{\bg}{\mathbf{g}}

\newcommand{\chem}{{\bm \mu}}

\newcommand{\fc}{\mathfrak{c}}

\renewcommand{\citenumfont}[1]{#1}
\renewcommand{\refdisp}[1]{Ref.~[\onlinecite{#1}]}
\renewcommand{\cite}[1]{~[\onlinecite{#1}]}

\usepackage{subfigure}
\usepackage{color}
\usepackage{bm}% bold math
\usepackage{alltt,dsfont}
\usepackage{appendix}
\usepackage{amsmath,amssymb}
\usepackage{hyperref}
\hypersetup{
    colorlinks,
    citecolor=red,
    filecolor=cyan,
    linkcolor=blue,
    urlcolor=magenta
}

\usepackage{atveryend}
\makeatletter
\let\origcitation\citation
\AtEndDocument{\def\mycites{}%
  \def\citation#1{\g@addto@macro\mycites{#1^^J}\origcitation{#1}}}
\AtVeryEndDocument{\newwrite\citeout\immediate\openout\citeout=\jobname.cit
  \immediate\write\citeout{\mycites}\immediate\closeout\citeout}
\makeatother

\begin{document}
\title{ Theory of anisotropic elastoresistivity of two-dimensional \\ extremely strongly correlated metals }
\author{ Michael Arciniaga$^1$, Peizhi Mai$^{2,1}$, B Sriram Shastry$^1$  }
\affiliation{$^1$Physics Department, University of California,  Santa Cruz, CA 95064, USA }
\affiliation{$^2$Center for Nanophase Materials Sciences, Oak Ridge National Laboratory, Oak Ridge, TN, 37831-6494, USA }
\date{\today}

\begin{abstract}
There is considerable recent interest in the phenomenon of anisotropic electroresistivity of correlated metals. While
some interesting work has been done on the iron-based superconducting systems, not much is known for the cuprate
materials. Here we study the anisotropy of elastoresistivity for cuprates in the normal state. We present theoretical
results for the effect of strain on resistivity, and additionally on the optical weight and local density of states. We
use the recently developed extremely strongly correlated Fermi liquid theory in two dimensions, which accounts
quantitatively for the unstrained resistivities for three families of single-layer cuprates. The strained hoppings of a
tight-binding model are roughly modeled analogously to strained transition metals. The strained resistivity for a
{two}-dimensional $t$-$t'$-$J$ model are then obtained, using the equations developed in recent work. Our quantitative
predictions for these quantities have the prospect of experimental tests in the near future, for strongly correlated
materials such as the {hole-doped and electron-doped high-$T_c$} materials.

\end{abstract}
\maketitle
%{\bf Keywords}
%\tJ Model, 

\section{Introduction \& Motivation}

Understanding the temperature and doping dependent electrical conductivity of very strongly correlated metals in {two}
dimensions (2D) is a very important problem in condensed matter physics. Recent interest in {\it elastoresistivity},
i.e., the strain dependence of resistivity has been triggered by the discovery of strong nematicity in iron based
superconductors\cite{Fisher2012,Fisher2015,Fisher2016}. The nematic susceptibility is defined as 
\beq
\chi_{nem} =\lim_{\epsilon_{xx} \to 0} \frac{\rho'_{xx}-\rho_{xx}}{\rho_{xx} \epsilon_{xx}} \label{nematic-chi}
\eeq
where  $\rho'_{xx}(\rho_{xx})$ is the x-axis resistivity in presence (absence) of a small strain $\epsilon_{xx}$. The large magnitude of this dimensionless susceptibility ($|\chi_{nem}|$$\gssim$$200$), and the peak like features in its temperature dependence suggest enhanced nematic fluctuations in the pnictides. 

The situation for cuprates is less studied thus   motivating the present work.  The recently developed extremely correlated Fermi liquid theory (ECFL)\cite{ECFL}  accounts quantitatively  for the (unstrained) normal state resistivities of three families of single layer cuprates \cite{S-M-New,SP,PS}. This theory treats correlation effects within the  well-defined  $t$-$t'$-$J$ model. The model lacks any explicit mechanism to  drive large nematic fluctuations, but it is possible that these fluctuations are emergent.
It is thus natural to ask if the theory can provide a benchmark scale for elastoresistivity effects in cuprates, as well as  to examine if nematic fluctuation  are encouraged. Towards this goal
we present results for the anisotropic elastoresistivity in various geometries for cuprate materials in the normal state
within the extremely correlated Fermi liquid theory (ECFL)\cite{ECFL} as applied to the $t$-$t'$-$J$ model for
spin-$\frac{1}{2}$ electrons on a square lattice given by the Hamiltonian
{
\beq
H= - \sum_{i j \si} t_{ij} \widetilde{C}_{i \si}^\dagger \widetilde{C}_{j \si} - \chem \hat{N} + \frac{1}{2}\sum_{i j}J_{ij} \left( \vec{S}_i.\vec{S}_j - \frac{1}{4} n_i n_j \right).  \label{Ham-0}
\eeq 
}
{Here $t_{ij} =t (t')$ for nearest (next-nearest) neighbour sites $ij$ and is zero otherwise on the square
lattice\cite{Http}}, $\hat{N}$ is the number operator, $\widetilde{C}_{i \si} = P_G C_{i \si} P_G$ and  $P_G$ is the
Gutzwiller projection operator {which projects out the doubly occupied states}. {Also the super-exchange $J_{ij}=J$ when acting on nearest neighbour sites and is zero
otherwise.} The other symbols have their usual meaning.

While the ECFL theory accounts for the variation  of resistivity with a change of hopping parameters, we need another piece of information to calculate elastoresistivity. {That is} a solution {to} the independent problem of describing the effects of strain on the hopping parameters of the
 underlying tight-binding model. In cuprates the $t$-$t'$-$J$ model arises as an effective low energy model from
 downfolding from a three band (or in general multi-band) description obtained within band structure
 calculations\cite{downfolding,downfolding1,downfolding2}. This procedure is not unique since the extent of
 correlations included in the band structure can differ {among} different calculations. We take the practical view that
 the hopping parameters can be chosen to depend parametrically on the distance between atoms, in parallel to the
 treatment of volume effects in transition metals by V. Heine\cite{Heine}. Thus in our approach, {a small} strain can
 be parametrized through a single variable $\alpha$  relating the hopping to  the separation $R$ via the relation 
\beq  t(R)\sim\frac{A}{R^\alpha}. \label{t-versus-R} \eeq
From tight binding theory  $\alpha =  l_1 + l_2 + 1$, where $l_1,l_2$ are the angular momenta of the overlapping orbitals \cite{Heine}. Within this scheme we expect that
 compression enhances overlap and hence the magnitude of hopping, and conversely stretching reduces overlap.  Excluding very strong multi-band effects  we may take  $\alpha\in\{2,5\}$ for cuprates. The single parameter needed for our purpose is $\alpha$, since $A$ is reabsorbed in the unstrained hopping. We further suggest that one may more realistically {estimate} this single
parameter $\alpha$ by  measuring other $\alpha$ dependent variation of  physical variables  with strain,  as described below. 

This modeling neglects the possible 3-dimensional effects, where the c-axis propagation could in certain situations  influence the 2-dimensional bands indirectly. Also cuprates with many layers per unit cell may have more complex dependence on strain as compared to single layer systems.
{Despite} the above caveats in place, it {is still} worthwhile to study the model \disp{Ham-0} together with the relation \disp{t-versus-R} for understanding the elastoresistivity of single layer cuprates.

 The problem of (unstrained) normal state resistivity has been explored in various experiments\cite{Ando, NCCO, Greene}
 on different materials over last few decades. Experiments reveal interesting and challenging transport regimes, termed
 the {\it strange metal} and the {\it bad metal} regime\cite{StrangeMetal}, whose existence is inexplicable within the standard Fermi
 liquid theory of metals. These results have attracted several numerical studies using the techniques of dynamical mean
 field theory\cite{DMFT,HFL,badmetal}, determinant quantum Monte-Carlo method\cite{dqmc,edwin} and dynamical cluster
 approximation\cite{dca,dcar} etc. These studies indicate that the unusual regimes are indicative of very strong
 correlations of the Mott-Hubbard variety.

Despite the numerical progress, few analytical techniques are available to extract the low temperature transport
behavior, and thus better understand the various regimes. This is due to the inherent difficulties of treating strong
correlations, i.e., physics beyond the scope of perturbation theory. Recently, the extremely correlated Fermi liquid
theory (ECFL)\cite{ECFL,ECFL2,Sriram-Edward} has been developed by Shastry and coworkers. This theory consists of a
basic reformulation of strong correlation physics, and its many applications have been reported for the \tJ model
in dimensions d=1,2,$\infty$. This is a minimal and fundamental model to describe extreme correlations. The ECFL
theory leads to encouraging results which are in close accord with experiments such as spectral line shape in
angle-resolved photoemission spectroscopy (ARPES)\cite{ARPES1,ARPES2,ARPES3,ARPES4,ARPES5,Gweon,SP,PS}, Raman
susceptibility\cite{raman,Koitzsch}, and particularly, resistivity\cite{Sriram-Edward,WXD,SP,PS}. A recent work \cite{S-M-New} shows that the ECFL theory gives
a quantitatively consistent  account  of the $T$ and density dependence of the resistivity  for single layer hole-doped and
electron-doped correlated materials. Here we explore the strain dependence of the resistivity within the same scheme.

In the ECFL theory, the resistivity arises from (umklapp-type) inelastic scattering between strongly correlated
electrons. Here the hopping amplitudes of electrons play a dual role. The first one, that of propagating the fragile
quasiparticles, is standard in all electronic systems. They  provide a simple model for the band structure. Additionally, for very strong
correlations the ECFL theory shows that the hopping parameters are also involved in the scattering of quasiparticles
off each other \cite{comment-1}. A surprisingly low
characteristic temperature scale\cite{WXD,PS} emerges from the strong correlations, {above which} the resistivity
crosses over from Fermi liquid type i.e. $\rho\sim T^2$ behavior, to an almost linear type i.e. $\rho\sim
T$ behavior\cite{SP,PS,Ando}.

From the above we argue that strain effects could provide a test of the underlying mechanism for resistivity within the
ECFL theory to include strain dependence. Experiments probing these strain effects are likely in the near future, thus
enabling an important test of the theory. For the purpose of independently estimating the strain-hopping parameter $\alpha$ in \disp{t-versus-R}, we have
identified two experimentally accessible variables. Firstly we study the integrated weight of the anisotropic
electrical optical conductivity, i.e., the f-sum rule weight, accessible in optical experiments\cite{optical1,optical2}.
Secondly we study the local density of states (LDOS), measurable through scanning tunneling microscopy
(STM)\cite{STM1,STM2,STM3,STM4,STM5}. The f-sum rule weight in tight binding systems is related to the expectation of
the kinetic energy, or hopping, and can be {obtained} from the Green's function. The LDOS can {also} be calculated from the
local Green's function easily.

The plan of the paper is as follows: In Sec. \ref{sec:MethodsParams} (A) we introduce the {\tJ} model and summarize the
second order ECFL equations and the corresponding Green's functions and self-energies. (B) We describe how to convert
the lattice constants and hopping parameters for a system under strain. (C) We outline the parameters for the program. In
Sec. \ref{sec:Results}, we present the detailed calculation for and results of (A) the resistivity, (B) the kinetic energy, and
(C) the LDOS and their associated susceptibilities with respect to strain. We provide a brief summary and discussion of
our results and future work in Sec. \ref{sec:Conclusion}.

\section{Methods \& Parameters} 
\label{sec:MethodsParams}

\subsection{The Model}
It has been argued that the {\tJ} model is key to describing the physics of high-$T_c$ superconducting
    materials\cite{Anderson}. This model is composed of two terms: $H_{tJ} = H_{t} + H_{J}$ where $H_{t}$ is derived by taking the
    infinite-U limit of the Hubbard model plus an additional term $H_J$ which introduces antiferromagnetic coupling. The
    general Hamiltonian \disp{Ham-0} can be rewritten in terms of the Hubbard $X$ operators \cite{ECFL} as
\begin{align} \label{eq:Hamiltonian}
    H_{t} &= -\sum_{ij\sigma} t_{ij} X^{\sigma0}_i X^{0\sigma}_{j} - \chem \sum_{i\sigma}X^{\sigma\sigma}_{i}\;, \\
    H_{J} &= \frac{1}{2}\sum_{ij\sigma}J_{ij}X^{\sigma\sigma}_i \nn \\
          &\;\;+ \frac{1}{4}\sum_{ij\sigma_1\sigma_2}J_{ij}
          \{X^{\sigma_1\sigma_2}_i X^{\sigma_2\sigma_1}_j 
          - X^{\sigma_1\sigma_1}_i X^{\sigma_2\sigma_2}_j \} \nn
\end{align}
Here $t_{ij}$ and $J_{ij}$ are already defined below Eq.~(\ref{Ham-0}). We present results for both vanishing and non-vanishing $J_{ij}$. The operator $X^{ab}_{i} = \ket{a}\bra{b}$ takes the electron at site i from the state $\ket{b}$ to the state $\ket{a}$
where $\ket{a}$ and $\ket{b}$ are one of the three allowed states: two occupied states $\ket{\uparrow}$,
$\ket{\downarrow}$, or the unoccupied state $\ket{0}$ --- {the appropriate $X$ operator referring to the doubly
    occupied state $\ket{\uparrow\downarrow}$ is excluded in both the Hamiltonian and state space.} {The $X$
    operator relates to the alternative representation used in Eq.~\ref{Ham-0} as follows: $X_{i}^{\sigma 0} \to
\widetilde{C}_{i\sigma}^{\dagger}$, $X_{i}^{\sigma 0}\to \widetilde{C}_{i\sigma}$ and
$\sum_{\sigma}X_{i}^{\sigma\sigma}\to n_{i}$.} 
\subsection{The ECFL Equations}
{In this section, we briefly introduce the ECFL equations for the \tJ model. More details can be found in
\refdisp{ECFL,ECFL2,Sriram-Edward,SP}.} In the ECFL theory, the one-electron Green's function $\G$ is found
using the Schwinger method\cite{Kadanoff} and in momentum space is factored as a product of an auxiliary
Green's function $\bg$ and a ``caparison'' function $\widetilde{\mu}$:
\begin{equation} \label{eq:product}
    \G(k) = \bg (k) \times \widetilde{\mu}(k)
\end{equation}
where $k \equiv (\vec{k}, i\omega_{k})$, and $\omega_{k} = (2k+1) \pi k_B T$ is the Fermionic Matsubara
frequency and subscript $k$ is an integer. The auxiliary
$\bg(k)$ plays the role of a Fermi-liquid type Green's function whose asymptotic behavior is $1/\omega$ as $\omega \to
\infty$, and $\widetilde{\mu}$ is an adaptive spectral weight that mediates between two conflicting
requirements\cite{ECFL2}: (1) the high frequency behavior of the non-canonical fermions and (2) the Luttinger-Ward
volume theorem at low frequencies.

The Schwinger equation of motion for the physical Green's function can be symbolically written as\cite{ECFL2}
\beq
\left( \GHI_{0} - \hat{X} - {Y_1}\right). ~\G = \delta \ ( \iden - {\gamma} ). \label{Min-1}
\eeq
where $\hat{X}$ represents a functional derivative and $Y_1$ describes a Hartree-type energy, i.e., $\G$
convoluted with hopping and exchange interactions. The left hand side of \disp{Min-1} is analogous to that of the
Schwinger-Dyson equation for Hubbard model\cite{SchwingerDyson}: $\left(  \GHI_{0} -  U \delta/{\delta{\V}} -  U  G \right). G = \delta \
\iden$. Observe on the right side of \disp{Min-1}, the essential difference is the $\gamma$ term which is proportional
to a local $\G$ and originates from the non-canonical algebra of creation and annihilation operators.  The non-canonical
nature of operators and the lack of an obvious small parameter for expansion present the main difficulties towards
solving this equation.

To tackle these difficulties, the ECFL theory inserts into \disp{Min-1} the $\lambda$ parameter
\beq
    \left(  \GHI_{0} -  \lambda \hat{X}-  \lambda {Y_1}\right). ~\G = \delta \  ( \iden  - \lambda {\gamma}  ). \label{Min-2}
\eeq  
where $\lambda \in [0,1]$ interpolates from a non-interacting to fully interacting system. This parameter plays a
parallel role to that of inverse spin parameter $1/2S$ in quantum magnets, where $S$ is the magnitude of the spin. Then we expand \disp{Min-2}
systematically with respect to $\lambda$ up to a finite order and at the end set $\lambda=1$ to recover the full \tJ
physics. The introduction of $\lambda$ bound to $[0,1]$ in ECFL makes it possible that a low-order expansion could be
enough to describe low-energy excitations in a large region of doping. This argument has been justified in one\cite{PSS}
and infinite\cite{Sriram-Edward} dimensions by benchmarking against exact numerical techniques and in two\cite{SP,PS}
dimensions by comparing well with experiments.

In the following, we use the minimal {version of} second order {(in $\lambda$)} ECFL equations\cite{SP}: {
\begin{align} 
    \widetilde{\mu}(k) &= 1 - \lambda \frac{n}{2} + {\lambda} \psi(k) \label{eq:ECFL1} \\
    \bg^{-1}(k) &= i\omega_{k} + \chem - \epsilon_{\vec{k}} + {\lambda} \frac{n}{2}\epsilon_{\vec{k}} - \lambda \phi(k)
    \label{eq:ECFL2}
\end{align}} 
where $\chem$ is the chemical potential (denoted in boldface) and $\epsilon_{\vec{k}}$ is the bare band energy found by taking the
Fourier transformation of the hopping parameter. The physical Green's function features two
self-energy terms: the usual Dyson-like self-energy denoted $\phi(k)$ in the denominator and a second self-energy in the
numerator $\psi(k)$. The self-energy $\phi(k)$ can conveniently be decomposed as follows: $\phi(k) = \chi(k) +
\epsilon_{\vec{k}}'\psi(k)$ where $\chi(k)$ denotes a self-energy part, $\epsilon'_{\vec{k}} =\epsilon_{\vec{k}} - u_0 /
2$ and $\psi(k)$ the second self-energy. Here $u_0$ acts as a Lagrange multiplier, enforcing the shift
invariance\cite{ECFL,ECFL2,SP} of the \tJ model at every order of $\lambda$. The two self-energies functions
    $\psi$ and $\chi$ expanded formally in $\lambda$ to second order approximation $\mathcal{O}(\lambda^2)$ are $\psi =
    \psi_{[0]} + \lambda \psi_{[1]} + \ldots$ and $\chi = \chi_{[0]} + \lambda\chi_{[1]} + \ldots $. The expression for these self-energies in
    the expansion are 
\bal \label{eq:selfenergy1}
\psi_{[0]}(k) = 0, \;\;\;\; {\chi_{[0]}(k) = -\sum_{p} \Big
(\epsilon'_{\vec{p}}+\frac{1}{2}J_{\vec{k}-\vec{p}} \Big )\bg(p)}
\eal
and 

\begin{align}\label{eq:selfenergy2}
    \psi_{[1]}(k) &= -\sum_{pq}\Big (\epsilon'_{\vec{p}} + \epsilon'_{\vec{q}} + J_{\vec{k}-\vec{p}} \Big ) \bg(p)\bg(q)\bg(p + q - k) \\
    \chi_{[1]}(k) &= -\sum_{pq} \big (\epsilon'_{\vec{p}} + \epsilon'_{\vec{q}} + J_{\vec{k}-\vec{q}} \big )
    \big (\epsilon'_{\vec{p}+\vec{q}-\vec{k}} + J_{\vec{k}-\vec{p}} \big )\nn \\ &\quad \times\bg(p)\bg(q)\bg(p + q - k) \nn \\
\end{align}

where $\sum_k \equiv \frac{k_B T}{N_s}\sum_{\vec{k},\omega_k}$ {and $J_{\vec{q}}$ is the Fourier transform of
$J_{ij}$~\cite{HJ}}. By setting $\lambda$ to $1$, the resulting expressions for the ECFL equations expanded to
$\mathcal{O}(\lambda^2)$ are
{
\begin{align}\label{second}
    \widetilde{\mu}(k) &= 1 - \frac{n}{2} + \psi(k) \\ 
    \bg^{-1}(k) &= i\omega_{k} + \chem - \epsilon_{\vec{k}} + \frac{n}{2}\epsilon_{\vec{k}} - \chi_{[0]}(k) \\  
                &\quad\quad-\chi_{[1]}(k) - \epsilon'_{\vec{p}} \psi_{[1]}(k) \nn 
\end{align}
}
We can verify that an arbitrary shift of { $\epsilon_{\vec{k}} \to  \epsilon_{\vec{k}} + {\bf c}_0 $} leaves the above expression
invariant by shifting {$\chem \to \chem + {\bf c}_0$ and $u_0 \to u_0 + 2{\bf c}_0$.} {In this sense, we may take $u_0$ as a second
chemical potential.}  We can determine the two chemical potentials
$\chem$ and $u_0$ by satisfying the following number sum rules
{
\begin{equation} \label{eq:numsumrules}
    \sum_k \bg(k)e^{i\omega_{k} 0^{+}} = \frac{n}{2} = \sum_k\G(k)e^{i\omega_{k} 0^{+}}\;,
\end{equation}
}
where $n$ is the particle density.
We find the spectral function $\rho_{\G}(k) =
-1/\pi \Im m \G(k)$ by analytically continuing (i.e. {$i\omega_{k} \to \omega + i\eta$}) and by solving
{Eq.~(\ref{eq:product}) and Eqs.~(\ref{eq:selfenergy1}-\ref{eq:numsumrules})} {iteratively}.
We remind the reader that the spectral function $\rho_{\G}(\vec{k},\omega)$ is referred to in most experimental literature by the
symbol $A(\vec{k},\omega)$.
We can recover the interacting Green's function from $\rho_{\G}$ using 
\begin{align}
    \G(\vec{k},i\omega_{k}) = \int_{-\infty}^{\infty}\frac{\rho_\G(\vec{k},\nu)}{i\omega_{k} - \nu}d\nu\;.
\end{align}

\subsection{Strain effects on hopping and exchange}

{\bf \S Converting lattice constant changes to hopping changes:} 
The t-t'-J model in {two} dimensions 
describes the hopping of electrons between copper atoms  in the 2-d plane.
In this model,
the hopping parameters with strain and without strain are denoted as 
\beq \label{eq:txtytd} 
\{t_x, t_y, t_d\} \to \{t, t, t'\}. 
\eeq
Thus under strain $t_x$ and $t_y$ refer to nearest neighbor hops along x and y axes, and $t_d$ is the second neighbor
hopping along the diagonal  of the square lattice. We start with the tetragonal symmetry case $t_x=t_y=t$ where there
are just two parameters $t,t'$.

At the level of a single bond between
two coppers, any generic hopping $t(R)$ for a bond with length $R$ can be represented by\cite{Heine}
\begin{equation}
t(R) \sim \frac{A}{R^\alpha} \label{t-versus-R-1}
\end{equation}
where $A$ is a constant. In the simplest cases, the exponent $\alpha$ is given by the angular momentum $l_1,l_2$ of the
relevant atomic shells of the two atoms by the formula 
\begin{equation} \label{alpha}
\alpha = l_1 + l_2 + 1\;.
\end{equation}

Thus for two copper atoms $l_1=l_2=2$ and hence we might expect
\begin{equation}
\alpha \sim 5,
\end{equation}
whereas for copper oxygen bonds $l_1=2, \; l_2=1$, therefore 
\begin{equation}
\alpha \sim 3\;.
\end{equation}
For the effective single band description of the cuprate materials, it is not entirely clear what value of $\alpha$ is
most appropriate. Comparisons with experiments might be the best way to decide on this question, when the results
become available. Until then we can bypass this issue by presenting the theoretical results in terms of $\frac{\delta
t}{t}$ rather than the strain itself. Towards this end \disp{t-versus-R-1} is a very 
useful result. We rewrite it as 
\begin{equation}
\frac{{\delta t(R)}}{t(R)}= - \alpha \; \frac{\delta R}{R}, \label{t-versus-R-2}
\end{equation} 
thus enabling us to convert a change of the lattice constant to that of the corresponding hopping, using only the value
of $t$ and $\alpha$. Throughout this paper we will refer to $\delta t / t$ as ``strain'' or with emphasis as ``hopping
strain'' in order to distinguish it from ``conventional strain'' $\delta R / R$.  Strain will always refer
to variations along the $x$-axis unless otherwise noted.

{\bf \S Geometrical aspects of the strain variation}

Our calculation studies a few variations of parameters. We start on a
lattice with tetragonal symmetry at $t \sim 5220$K ($0.45$eV), and we vary $t'$ to capture both electron-doped ($t'>0$) and hole-doped ($t'<0$)
cuprates. The magnitude of $t$ is only a crude estimate, it is refined for different single layer cuprate systems in \refdisp{S-M-New}. 

 On the distorted lattice with orthorhombic symmetry and
lattice constants $a$ and $b$, the three distances of interest (two sets of
nearest neighbors and one set of second neighbors) are 
\begin{equation} 
a, \; b,\;\; \rho= \sqrt{a^2+b^2}. 
\end{equation}
For the tetragonal case we refer to the undistorted lattice parameter as
$a_0$, thus
 $a = b = a_0$, $\rho = \sqrt{2} a_0$. We next study the effect of stretching ($\delta a>0$) or compressing ($\delta a<0$) the $x$-axis lattice
constant, leaving the
$y$-axis unchanged. The changes in the lattice constants then read as 
\begin{equation}
a \to a_0 + \delta a; \;\; b\to a_0; \;\; \rho \to \sqrt{2} a_0 + \frac{ \delta a}{\sqrt{2}}\;. \label{eq:lattice-constants}
\end{equation}
We denote the strain in the x-direction as
\begin{equation}
\epsilon_{xx}=  \frac{\delta a}{a_0}.
\end{equation}
In terms of the strain, we can rewrite the distances to neighbors as
\begin{equation}
a = a_0 (1 + \epsilon_{xx}), \;\;  b = a_0, \;\; \rho= \sqrt{2} a_0 \left(1 + \frac{\epsilon_{xx}}{2} \right) \;, \label{stretches}
\end{equation}
so that $\epsilon_{xx}>0$ is regarded as stretching and $\epsilon_{xx}<0$ as compression. The single particle 
(tight-binding) energies for the distorted lattice are given by 
$$
    \epsilon_{\vec{k}} = -2t_{x}\cos(k_{x}a)-2t_{y} \cos(k_{y}b)  
                -4t_{d} \cos(k_{x}a)\cos(k_{y}b). 
$$
In terms of the band parameters of the unstrained system $t$ and $t'$, we can write the anisotropic band parameters as
\beq \label{eq:hopparams}
t_{x} = (1 -  \alpha \; \epsilon_{xx}) \; t , \;\;
t_{y} = t , \;\;
t_{d} = \Big (1 - \alpha \; \frac{\epsilon_{xx}}{2} \Big ) \; t' \;, 
\eeq
where the factor of $\frac{1}{2}$ for $t_d$ comes about due to a shorter stretching of $\rho$ as in \disp{stretches}. Their strain variations  are denoted by
\beq
\frac{\delta t_{x}}{t_x} \equiv \frac{\delta t}{t} = - \alpha \epsilon_{xx}, \;\;\frac{\delta t_{y}}{t_y}=0, \;\;\frac{\delta t_{d}}{t_d}= -\frac{1}{2} \alpha \epsilon_{xx}. \label{delta-t-epsilon}
\eeq
These formulas relate the change in hopping to the physical strain, and thus involve the parameter $\alpha$ which is
somewhat uncertain. For that reason, we actually vary $\frac{\delta t}{t}$ in this study. We also go beyond the linear
response regime, i.e., we use larger values of $\frac{\delta t}{t}$  than those attainable in the laboratory. In such a case we set $\frac{\delta t_{d}}{t_d}=\frac{\delta t}{2 t}$. To summarize the sign convention used in this work, 
\begin{eqnarray}  
&\mbox{compress:}  \;\;&\frac{\delta t}{t}>0, \;\; \epsilon_{xx}<0 \nn \\
&\mbox{stretch:}  &\frac{\delta t}{t}<0,  \;\; \epsilon_{xx}>0. \label{signs}
\end{eqnarray}

{
{\bf \S Converting hopping changes into exchange changes:}
{In this model, the super-exchange} interaction maps to hopping as follows: $J = t^2 / U$ where $U$ is the on site energy of
the Hubbard model. As we vary the hopping parameter, we find $\delta J= 2 (\delta t / t) J$ since $U$ does not vary with
strain. In this model the first neighbor exchange parameters with and without strain, similar to Eq.~\ref{eq:txtytd}, are denoted as 
\begin{equation}
    \{J_x ,J_y\} \to \{J,J\}\;,
\end{equation}
where $J_x$ and $J_y$ refer to the first neighbor exchange interactions along the x and y axes. In terms of hopping
changes we can rewrite the exchange parameters as 
\begin{equation} \label{eq:exchparams}
    J_{x} = \Big (1 + 2 \frac{\delta t_{xx}}{t_{xx}}\Big ) \; J , \;\;
    J_{y} = J\;.  \;\;
\end{equation}
}

\subsection{Parameters in the program}
The model considered applies to several classes of materials, such as the cuprates, the sodium cobaltates, and
presumably also to the iron arsenide superconductors. We shall restrict our discussion to the cuprates where the
parameters are fairly well agreed upon in the community\cite{Anderson,Ogata,S-M-New}. 

In this calculation, we set $t=1$ as our energy scale and we allow $t'/t$ to vary between $-0.4$ and $0.4$, to cover the
full range of cuprate materials. The hopping strain $\delta t / t$ is varied from $-0.15$ to $0.15$. {The exchange
parameter $J$ is set to zero {except} where otherwise noted.} We convert the
energy to physical units by setting $t=0.45$eV, and hence the bandwidth is $W=8t=3.6$eV.
If one wants to make a different choice for $t$, this  can be done  by rescaling the energies and T's by the same scaling factor.

We focus on the optimal doping case $\delta = 0.15$ for cuprate materials\cite{materials}. Here $\delta$ refers to the hole
doping and relates to the particle density as follows $\delta \equiv (1 - n)$. The temperature range is set to $T \in
[37,450]$K. Lower temperatures than this lie outside the range of convergence
for {the current} scheme. For the interacting system we solve the ECFL {equations
(\ref{eq:selfenergy1}-\ref{eq:numsumrules})} iteratively on a real frequency grid of size $N_\omega = 2^{14}$ {within the range [-2.5W,2.5W], where W is the  bare bandwidth,} and a lattice $L
\times L$ with $L = 61, 79, 135$. The scale of the frequency grid is tuned to capture the low-$T$ physics. A frequency
grid of size $N_\omega = 2^{16}$ only slightly improves our results at much larger computational costs. We primarily use an $L
> 61$ for $t' > 0$ at low temperatures (i.e., $T < 100$K) in order to get sufficient resolution to converge electrical
resistivity calculation. The need for a high resolution lattice at low temperatures is a product of the spectral
function which features higher, sharper peaks for $t'>0$, to which the resistivity calculation is
sensitive\cite{SP}, i.e., a larger grid is required to settle the unphysical oscillations in the resistivity calculation. For the
non-interacting system we compute LDOS using a system of size $N_\omega=2^{12}$ and $L=271$.

\section{RESULTS}
\label{sec:Results}
Here we present the effects of strain along the x-axis on electrical resistivity, kinetic energy and
LDOS and their associated susceptibilities in response to a compressive ($\delta t / t > 0$) and tensile ($\delta t / t
< 0$) hopping strain. 

\subsection{Resistivity for an x-axis strain:}

We now study the response of electrical resistivity $\rho_{\alpha}$ characterized by electron-electron
scattering\cite{SP} in the presence of a strain. We use the bubble approximation, factoring the current
correlator as $\langle J(t) J(0) \rangle \sim \sum_{k} v_{\vec{k}}^{2}\G^2(k)$ with suitable vertices $v_{\vec{k}}$ and
dressed Green's function $\G$, to
    compute the conductivity $\sigma_{\alpha}$. Our picture of a quasi-2D metal consists of well separated Cu-O planes and hence each plane can be characterized
using the 2D $t$-$J$ model. The weak k-dependence of the self-energy as seen in Fig.~3 of
    Ref.~\cite{PS} diminishes the significance of vertex corrections. In fact the self-energy is completely
    k-independent in the d=$\infty$ limit and studies in this limit\cite{WXD} have successfully implemented the bubble
    approximation while completely ignoring vertex corrections. We shall calculate and quote the following objects denoting  the  irreducible representations   of the $D_{4h}$ point group by the standard names \cite{Tinkham,Hamermesh,LL,Fisher2015}
\begin{itemize}
\item $\rho_{xx}'(T)$ the strained version of resistivity along x-axis. 
\item $\rho_{yy}'(T)$ the strained version of resistivity along y-axis. 
\item $\rho_{xx}$ without a prime refers to the tetragonal result{, which is} the same as $\rho_{yy}$.
\item $XX$ component variations: $$-(\rho'_{xx} - \rho_{xx}) / (\rho_{xx} \delta t / t) \; \text{vs} \; T$$ 
\item $YY$ component variations: $$-(\rho'_{yy} - \rho_{yy}) / (\rho_{xx} \delta t / t) \; \text{vs} \; T$$ 
\item $A_{1g}$ symmetry variations: $$-\frac{\rho'_{xx} + \rho'_{yy} - 2\rho_{xx}}{2\rho_{xx} \delta t / t} \; \text{vs} \; T$$  
\item $B_{1g}$ symmetry variations:
    $$-\frac{\rho'_{xx} - \rho'_{yy}}{\rho_{xx} \delta t / t} \; \text{vs} \; T$$ 
\end{itemize}
Of special interest are the $\rho'_{xx} + \rho'_{yy}$ response which corresponds to the $A_{1g}$ irreducible
representation (irrep) and $\rho'_{xx}-\rho'_{yy}$ response, corresponding to the $B_{1g}$ irrep.

{\bf \S Computation of the anisotropic resistivity}

\begin{figure*}
    \includegraphics[width=\textwidth]{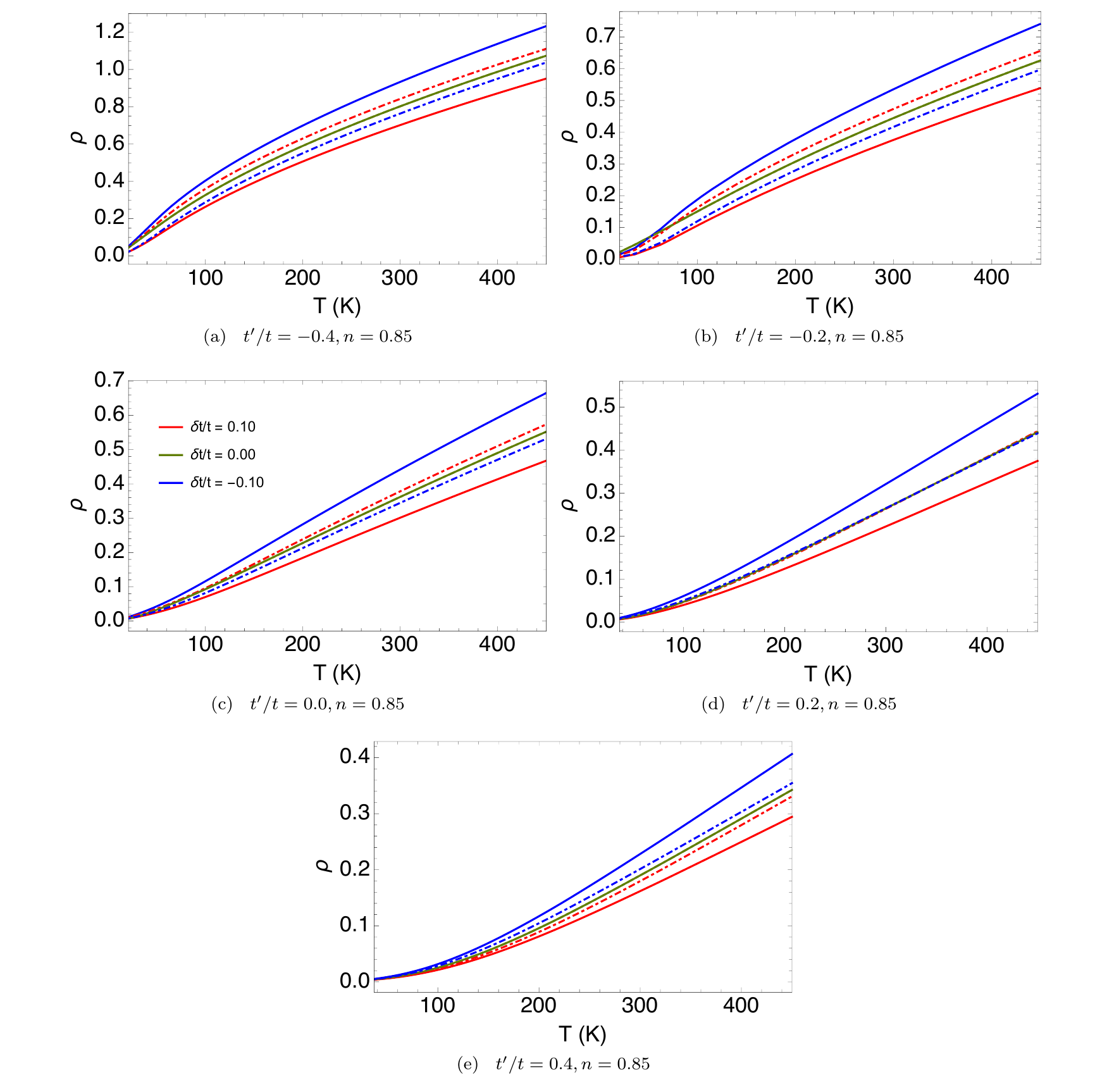}
    \caption{ The longitudinal and transverse resistivities, {$\rho_{xx}$} (solid) and {$\rho_{yy}$} (dashed)
        {respectively, versus temperature} at filling $n=0.85$, for various $t'$, and at  representative strains. While
        the green curves are for the unstrained case, the red curves correspond to a compressive strain $\delta
        t/t=0.10$ (i.e.  $\epsilon_{xx}\sim-.02$), and the blue curves correspond to a tensile strain $\delta t/t=-0.10$
        (i.e. $\epsilon_{xx}\sim.02$), if we take  $\alpha\sim 5$. All figures share the legend. The resistivity in
        physical units can be found by ${\rho}_{physical} = \rho
     \times \rho_0$, where $\rho_0=\fc_0 h/e^2 \sim 1.17$m$\Omega$cm.} \label{fig:rhoxxrhoyy}
\end{figure*}

To find the anisotropic resistivity, we compute the dimensionless conductivity\cite{SP} for the anisotropic case 
\begin{align} 
    \sigma_{xx} &= \langle \Upsilon_{\vec{k}} (\hbar v^{x}_{\vec{k}})^2 / (ab) \rangle_{k} \;, \\
    \sigma_{yy} &= \langle \Upsilon_{\vec{k}} (\hbar v^{y}_{\vec{k}})^2 / (ab) \rangle_{k}
\label{eq:sigmas}
\end{align}
where $\langle A \rangle_{k} = \frac{1}{N_s}\sum_{\vec{k}} A$, $N_s = L \times L$ and
\beq 
\Upsilon_{\vec{k}} = (2\pi)^{2} \int^{\infty}_{-\infty} d\omega (-\partial f / \partial \omega)
\rho^{2}_{\G}(\vec{k},\omega)
\label{eq:Upsilon}
\eeq
where $f(\omega) \equiv 1 / (1 + \exp(\beta\omega))$ is the Fermi function, $\rho_\G(k)$ is the spectral function from
from ECFL theory up to $\mathcal{O}(\lambda^2)$, and $v^{x}_{\vec{k}}$, $v^{y}_{\vec{k}}$ are the bare vertices, which are defined as
\barray 
v^{x}_{\vec{k}} &=& 
\frac{1}{\hbar}\frac{\partial \epsilon_k}{\partial k_{x}} = 
\frac{a}{\hbar}\frac{\partial \epsilon_k}{\partial k_{1}}\;, \\
v^{y}_{\vec{k}} &=& 
\frac{1}{\hbar}\frac{\partial \epsilon_k}{\partial k_{y}} = 
\frac{b}{\hbar}\frac{\partial \epsilon_k}{\partial k_{2}}
\earray
where $k_{1}= k_{x}a$ and $k_{2} = k_{y}b$ denote the components of the dimensionless momenta. Inserting the dimensionless momenta into
\disp{eq:sigmas}, we obtain
\begin{align} 
    \sigma_{xx} &= \Big \langle \Upsilon_{\vec{k}} \bigg (\frac{d \epsilon_{\vec{k}}}{d k_{1}} \bigg )^2 (a / b) \Big
        \rangle_{k} \;, \label{eq:modsigmas1} \\
    \sigma_{yy} &= \Big \langle \Upsilon_{\vec{k}} \bigg (\frac{d \epsilon_{\vec{k}}}{d k_{2}} \bigg )^2 (b / a)
        \Big \rangle_{k} \label{eq:modsigmas2}
\end{align}
for the dimensionless conductivity. The corresponding dimensionless resistivities are
$\rho_{xx} = 1 / \sigma_{xx}$ and $\rho_{yy} = 1 / \sigma_{yy}$.

The electrical resistivity can be converted to physical units as follows: ${\rho}_{physical, \alpha}=\rho_{\alpha}
\times \rho_0$ where $\rho_0 = \fc_0 h / e^2 (\sim 1.171$ m$\Omega$cm) sets the scale for the resistivity, and
$\alpha = xx$ describes the longitudinal (i.e., current $\parallel \epsilon_{xx}$) resistivity and $yy$ describes the transverse
(i.e., current $\perp \epsilon_{xx}$) resistivity. Here $\fc_0 \sim 6.645$\AA\ is the typical separation between parallel Cu-O
planes\cite{Distance,S-M-New}. In order to estimate the magnitude of the inelastic scattering, we can relate the
dimensionless resistivity to $\langle k_F \rangle \ell$ as follows $\langle k_F \rangle \ell = 1 / \rho_{\alpha}$ as
argued in Refs.~\cite{Ando,Ando2} for quasi-2D materials, where $\langle k_F \rangle$ is an (angle averaged)
effective Fermi momentum and $\ell$ is the mean-free-path. Hence we expect $\rho_{\alpha}/\rho_0 < 1$ in a good metal.

\subsubsection{The raw resistivities}
%\Mike{ In strongly correlated materials, it well known that Gutzwiller correlations leads to a Fermi liquid regime,
%followed by a strange-metal regime, a bad-metal regime, and finally a high-T regime with three crossover temperatures as
%illustrated in Fig.~1 of Ref.~\cite{WXD}} 
We first present the effects of hopping strain $\delta t / t$ on resistivity. In Fig.~\ref{fig:rhoxxrhoyy}, we study
the anisotropy of the raw dimensionless resistivity over a broad range of temperatures at the optimal density $n=0.85$.
Fig.~\ref{fig:rhoxxrhoyy} displays the longitudinal resistivity {$\rho_{xx}$} (solid) and the transverse resistivity
$\rho_{yy}$ (dashed) for a compressive strain (red) and tensile strain (blue) in comparison to the unstrained tetragonal
system (green). Here we used a representative magnitude of compressive strain $\delta t/t=0.10$ (i.e.
$\epsilon_{xx}\sim-.02$). We observe that longitudinal resistivity under a compressive strain ($\delta t / t > 0$) is
reduced, and conversely, under a tensile strain ($\delta t / t < 0$) it is enhanced across the displayed temperature
range for all $t'$. The response for transverse resistivity is less than the longitudinal one in magnitude. An
{interesting new feature} lies in the $t'$ dependence, we note that magnitude and sign of the change in transverse
resistivity is controlled by $t'$, e.g., for $t'=0.2t$ the resistivity is almost unchanged for all strains.

These behaviors can be understood qualitatively in the following ways. First, let us look at the simplest case with
$t'=0$ as in Fig.~\ref{fig:rhoxxrhoyy} panel (c). When the system is compressed in the x-axis, the hopping $t_x$ rises
according to \disp{eq:hopparams} and so does the conductivity along the same direction, and vise versa. Hence, the
longitudinal resistivity gets suppressed (enhanced) under compressive (tensile) strains. One can also consider isolating
the strain-induced effects in Eqs.~(\ref{eq:modsigmas1}) and (\ref{eq:modsigmas2}) from the band structure, contained in
$v_{\vec{k}}^{\alpha}$, and from the spectral function $\rho_\G$, which accounts for the influence of the Gutzwiller
correlations on resistivity. (Changes in the resistivity due to variation of the {explicit} lattice constants are small.) When we
exert a compressive strain, this produces additive changes to the longitudinal resistivity due to in equal parts (1)
changes in vertex and (2) $T$-dependent changes in spectral function, both arising from the enhancement of $t_x$.
Whereas for the transverse resistivity the hopping parameter $t_y$ is unchanged and hence changes to resistivity from the
band structure become less important and as a result the transverse resistivity is dominated by strain-induced effects
on the spectral function. For this reason, the transverse response to compressive strain is generally smaller in
magnitude than the longitudinal response and likewise for a tensile strain both shown in panel (c). We also find that
the transverse strain response has a different sign than the longitudinal one when there is no second neighbor hopping.

%These behaviors can be understood qualitatively in the following ways. First, let's look at the simplest case with $t'=0$ in Fig.~\ref{fig:rhoxxrhoyy} panel c. When the system is compressed in the x-axis, the hopping $t_x$ in the longitudinal direction rises and so does the conductivity along the same direction, and vise versa. Hence, the longitudinal resistivity gets suppressed (enhanced) under compressive (tensile) strain. To understand the transverse resistivity $\rho'_{yy}$ in the compressive case, one can consider an equivalent process, namely, starting from a smaller square lattice (with lattice constant $a$ and resistivity $\rho_a$) and exerting a tensile strain along the y axis so that the y-axis lattice constant becomes $a_0$. Then we can study $\rho'_{yy}$ as a longitudinal resistivity for the smaller square lattice. Based on our previous analysis, $\rho_{yy}$ should be larger than $\rho_{a}$ by a similar amount as $\rho_{xx}-\rho'_{xx}$ in the tensile case. Note that $\rho_{i}$ is smaller than $\rho'_{xx}$, which is our actual unstrained resistivity, because it has a smaller lattice constant than the unstrained case. Hence, $\rho_{yy}$ is less larger from $\rho'_{xx}$ than it is from $\rho_{i}$ due to this "counter" effect, likewise for the tensile case, both shown in panel c. For this reason, the transverse response is generally smaller in magnitude than the longitudinal response.

Now let us turn on $t'$. According to \disp{eq:hopparams}, the strain has a longitudinal-like effect, only smaller, on
the magnitude of the second neighbor hopping. Turning on a positive $t'$ strengthens longitudinal response and
``counters'' the transverse response from $t_y$ hopping. Therefore we see that the longitudinal curves depart further
from the unstrained one in panel (d) and (e), and it also explains why the transverse change almost vanish for $t'=0.2$
in panel (d) and switch to the same sign as the longitudinal one for $t'=0.4$ in panel (e). Likewise turning on a
negative $t'$ weakens the longitudinal response and enhances the transverse response, so that the longitudinal response
gets smaller in panel (a) and (b) while the transverse shifts more explicitly to the same side as $t'=0$.
%To understand the sign change it is helpful to recognize that $t'$ controls the sign of the strain-induced response of the spectral
%function and hence the sign of the transverse resistivity response since the role of $v_k^{\alpha}$ is negligible. That
%explains why the transverse change almost vanishes for $t'=0.2t$ in panel (d) and switches to the same sign as the
%longitudinal one for $t'=0.4t$ in panel (e). 
Further analysis of these effects can be found in the Supplementary Material (SM) \cite{SM}.

%\Peizhi{\sout{Now let's turn on $t'$. According to \disp{eq:hopparams}, when $t'$ and $t$ share the same sign, the $t_d$ term has a similar but less effect than the longitudinal $t_x$. Therefore we find the longitudinal response becomes larger with increasing $t'$ in panel d and e. From the analysis above for $t'=0$ case, the $t_y$ effect on resistivity has a opposite sign to the $t_x$ term. Turning on a positive $t'$ would weaken the transverse response and eventually force it to change sign. That explains the transverse change almost vanish for $t'=0.2$ in panel d and switch to the same sign as the longitudinal one for $t'=-0.4$ in panel e. Similarly, when $t'$ is negative, it counters the effect from $t_x$ term and adds to that from the $t_y$ term. Therefore, the longitudinal response gets slightly smaller in panel (a) and (b) while the transverse change becomes more explicit in the same side as in $t'=0$.}}

Next we discuss the how strain affects the effective interaction and the characteristic temperature scale. We mainly use
the longitudinal resistivity in this discussion because the longitudinal response is more explicit. In our recent
work\cite{PS,SP}, a significant finding was the $t'$ dependence of the curvature of the $\rho$ - $T$ lines. We observe
that this $t'$-dependent curvature persists under strain, i.e., the curvature changes from positive (concave up like
$+T^2$) to negative (convex up like $-T^2$) as $t'/t$ is varied upward. 

Recall that strain is effectively a small change in the hopping parameter, so we ought to expect strain to change the
$t'$ dependence of the curvature only quantitatively but not qualitatively.  Phenomenologically, varying $t'$ signals a
change in the effective Fermi temperature scale $T_{FL}$ where for $T < T_{FL}$ the system is in the Fermi liquid regime
$\rho \propto T^2$ and hence has a positive curvature. Moreover, as we decrease $t'$ from positive to negative, the
Fermi liquid temperature regime is compressed into a smaller temperature regime down to temperatures where resistivity
is usually hidden by the superconducting state. We want to focus on the crossover between Fermi liquid and strange metal which is covered by the following empirical  relation
\begin{equation} 
    \rho \sim C\; \frac{T^2}{T_{FL}+T}\;. 
\end{equation}
Here $C$ is a constant that defines the slope of linear regime and $T_{FL}$
marks the crossover from the Fermi-liquid regime. For example when $t'=-0.2t$ as found in typical hole-doped
cuprates\cite{holedoped}, we observe that a compressive strain extends the Fermi-liquid regime for the longitudinal
resistivity, and flipping the strain reduces the Fermi-liquid regime. Qualitatively speaking, a compressive strain
enhances the longitudinal hopping so that the effective interaction reduces relatively to the hopping. Likewise, a
tensile strain increases the effective interaction in the unit of longitudinal hopping and suppress the Fermi liquid
temperature scale. Besides, we observe that a compressive strain suppresses the linear constant C while a tensile
strain enhances it, as shown more obviously in \figdisp{fig:susrhoxxrhoyynd085}. That can be verified in the experiment
by measuring the slope of $\rho$ - $T$ for a {\it strange metal} under strain.
    
 %   \Mike{Furthermore, we can quantitatively compare $T_{FL}$ between the two directions with by
  %  examining the ratio of first and second neighbor hopping parameters $t_d/_x$ and $t_d/t_y$, i.e., increasing the
%ratio from negative to positive increases the temperature range of Fermi-liquid-like behavior.} 

%\sout{The difference in magnitude of the response between the longitudinal and transverse resistivity under compression is characterized by the relation $t_d / t_x > t_d / t_y$ and hence the longitudinal resistivity is more Fermi-liquid-like. Conversely, we see that a compressive strain reduces the Fermi-liquid temperature scale for the transverse resistivity sending the $\rho_{yy}$-$T$ curve to the linear regime more quickly, and enlarges the resistivity at high temperatures indicating an enhanced linear constant, C. }

%\sout{Although not shown in \cref{fig:rhoxxrhoyy}, we also note that increasing the magnitude of the compressive strain extends the Fermi-liquid regime of the longitudinal resistivity, all of which is consistent with the above relation.}

\subsubsection{Susceptibilities for anisotropic resistivities}

It has been argued\cite{Fisher2012} that cuprates are candidates for an electron nematic phase, in which nematic order
might coexist with high temperature superconductivity, that is, the electronic system breaks a discrete rotational
symmetry while leaving the translational symmetry intact. Here the normalized resistivity response plays the role of the
order parameter in the phase transition. Since it is possible to experimentally identify continuous phases transitions
through observation of a diverging thermodynamic susceptibility across a phase boundary this makes the temperature
profile of elastoresistance, i.e. normalized resistivity response with respect to an arbitrary strain, an interest
observable to explore. For that reason, we shall examine linear response function for the longitudinal and transverse
components of the elastroresistivity tensor constructed
in terms of the hopping strain as:
\begin{eqnarray}
    \chi_{XX} \equiv -\Big (\frac{\rho'_{xx}-\rho_{xx}}{\rho_{xx}} \Big ) 
    \Big / \Big (\frac{\delta t}{t} \Big ) \;, \label{norm-chiXX} \\
    \chi_{YY} \equiv -\Big (\frac{\rho'_{yy}-\rho_{yy}}{\rho_{xx}} \Big )
    \Big / \Big (\frac{\delta t}{t} \Big ) \;, \label{norm-chiYY}
\end{eqnarray}
respectively.  The susceptibility as defined is positive if compression along x-axis leads to a reduction of the resistivity in the specified direction. We note the connection of these susceptibilities with the nematic susceptibility \disp{nematic-chi} on using \disp{delta-t-epsilon} as
\beq
\chi_{nem}= {\alpha} \lim_{\epsilon_{xx}\to0} \chi_{XX}.  \label{nematic-chi-2}
\eeq
 We compute the susceptibility  for small  values of strain $\delta t/t\gssim .05$. However, even these values of strain  pick up some   non-linear components of the response function. These are also of interest, and we comment on these below.

\begin{figure}[h]
    \includegraphics[height=0.69\textheight, width=0.90\columnwidth]{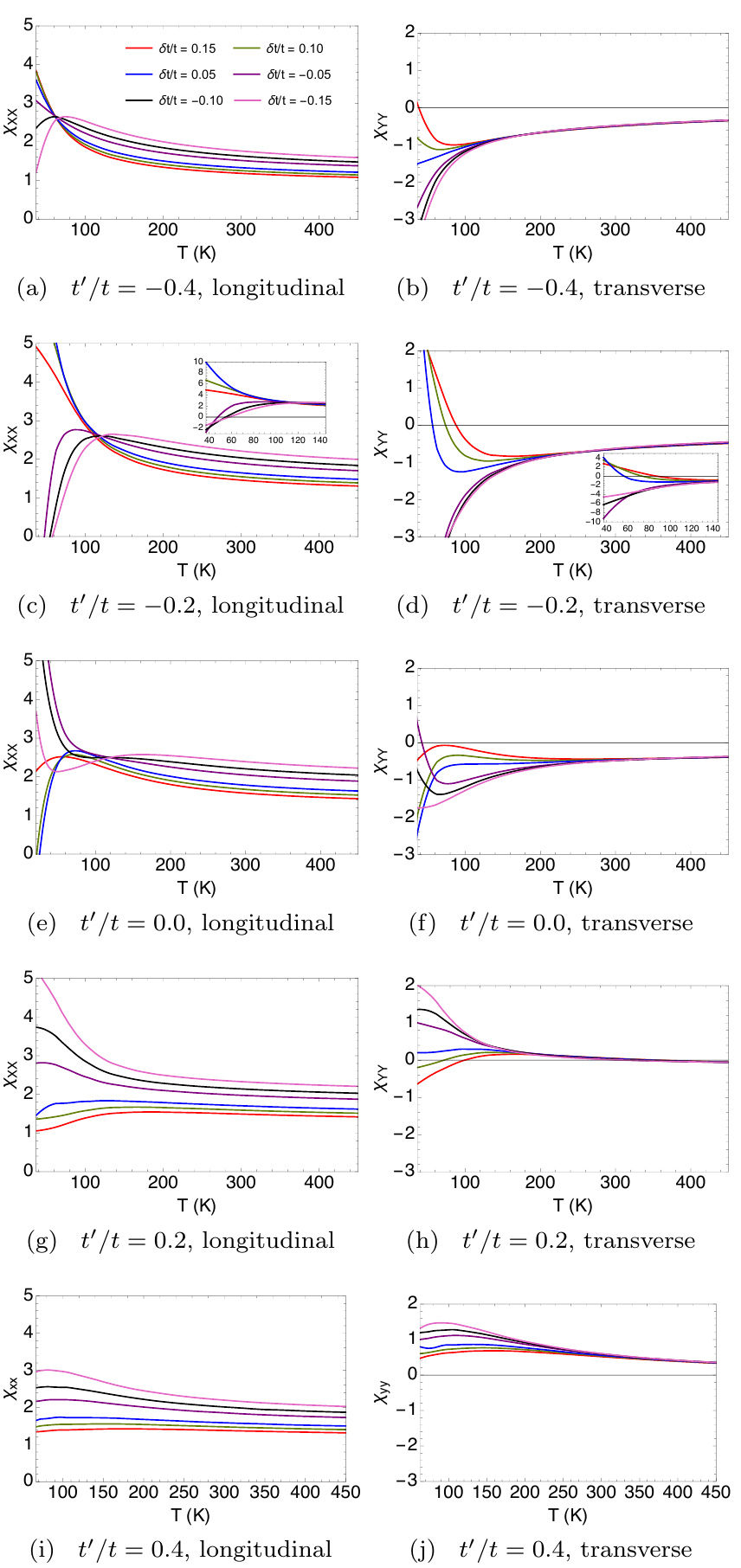}
    \caption{The longitudinal [\disp{norm-chiXX}] and transverse [\disp{norm-chiYY}] strain-resistivity
    susceptibilities versus temperature at filling $n=0.85$, for various $t'$ and $\delta t / t$. All figures share a
    legend. For various  $\delta t/t$,  the susceptibilities $\chi_{xx}$  for  $T\gssim 100$ approach  each other in  two sets, one for $\delta t/t>0$ and another slightly displaced set for  $\delta t/t<0$. They splay apart  at low T thus displaying strong non-linearity in the Fermi liquid regime. The susceptibilities $\chi_{yy}$ approach a single set for $T\gssim 100$  and splay apart for low T thus also displaying strong non-linearity in the Fermi liquid regime.    }
    \label{fig:susrhoxxrhoyynd085}
\end{figure}

The linear response function for strain-resistivity curves is plotted as a function of temperature in
\figdisp{fig:susrhoxxrhoyynd085} for the longitudinal and transverse components at optimal density $n=0.85$ for
various $t'$ and $\delta t / t$. Note that since the resistivity vanishes as $T \to 0$, there is an enhancement of the
normalized susceptibility at low-$T$.

In \figdisp{fig:susrhoxxrhoyynd085}, subfigures (a,c,e,g,i) we see that the linear response function for the
longitudinal resistivity $\chi_{XX}$ is mostly positive and shows non-linear (in $\delta t/t$) behavior at a fixed $T$
(as can be identified by the separation of the strain curves) with respect to strain across the entire temperature
range. This non-linearity will be measured directly in \figdisp{fig:susVSdtrhond085} for $t'=-0.2$. The response
function for $T \gssim 100$K is highly ordered in that varying the strain from positive (compressive) to negative
(tensile) increases the strength of the response function for all $t'$. Conversely as we cool the system, we observe
that strain dependence of the response function becomes increasingly non-linear, i.e., showing a wider separation
between strain curves, the forms of which are strongly $t'$ dependent. Now if we vary $t'$ to survey the range of
cuprate materials, we find at low-$T$ for hole-like ($t'<0$) materials a significant enhancement in and an inversion of
the strain dependence that is absent in electron-like ($t'>0$) materials, though for both material types the strength of
the response function remains approximately invariant at high-T.

     We  next discuss  the transverse linear response function $\chi_{YY}$ shown in  \figdisp{fig:susrhoxxrhoyynd085},
     subfigures (b,d,f,h,j).      This response is potentially interesting  since the affects of strain on the band
     structure are found to play a less significant role, hence the correlation effects dominate. We find that the
     features of transverse response function are different from that of the longitudinal response function mainly in
     two ways: (1) the $\chi_{YY}$ collapses at high-$T$, showing strong linearity with respect to the strain and (2) it
     changes sign from negative to positive as we vary $t'/t$ across 0.2 from below, consistent with
     \figdisp{fig:rhoxxrhoyy}.    Measurements confirming this linear behavior and sign change would be potentially
     interesting results.

    %\sout{If we vary the strain at low temperatures, we observe that strength of the
    %response function is highly $t'$ dependent, whereas at high-T the strength of the response function for given a
    %material is approximately independent of magnitude and direction of the strain, hence we say the response function
    %is linear at high-T.} 
    
    %\sout{We also find that the transverse response function at high-T changes sign from negative to positive as vary $t'$ from
    %hole-like to electron-like systems.}

     {

\subsubsection{Resistivity with non-zero $J$}

\begin{figure}[h]
    \includegraphics[width=0.99\columnwidth]{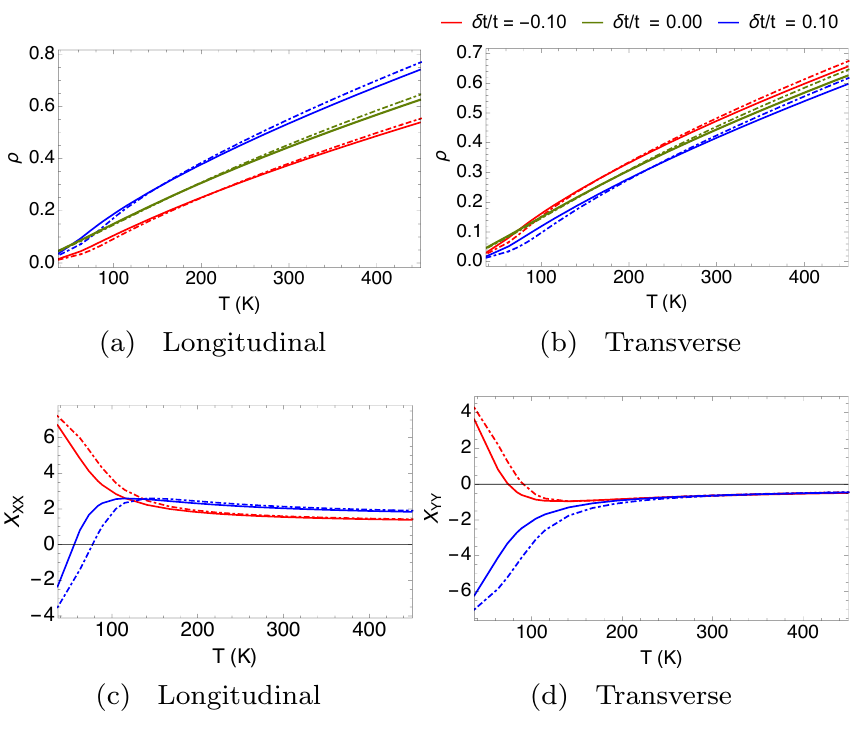}
    \caption{(a)-(b) The strain-induced resistivity at optimal filling $n=0.85$, $t'/t=-0.2$ for three
            representative strain types with exchange parameter
        $J=0.0$ (solid) and $J = 0.17$ (dashed). (c)-(d) The strain-resistivity susceptibility for the same parameter
set as above. All figures share a legend. We note that a non-vanishing $J$ enhances somewhat the magnitude of the susceptibilities in the low temperature Fermi liquid regime.}
    \label{fig:rhond085tpM020J017}
\end{figure}
In this section we examine the role of exchange parameter $J$ (nearest neighbor exchange energy) on resistivity and the susceptibilities, setting $J =
0.17t$ which is the typical value for LSCO cuprate materials \cite{Ogata}. We take $J = t^2 / U$ where $U$ is the on site energy of
the Hubbard model and $U$ does not vary with strain and hence $\delta J= 2 (\delta t / t) J$ \cite{HJ}. Now, if we turn
on the exchange parameter $J$, we find that at low temperatures the resistivity is reduced by the exchange energy and at
high temperatures the resistivity is slightly enhanced as seen in Fig.~\ref{fig:rhond085tpM020J017} panels (a) and (b).
In panels (c) and (d) we see the longitudinal and transverse susceptibility with exchange interaction {is further
enhanced} at low-temperatures whereas at higher temperatures the response is unchanged.
The $J$ effects are magnified in the low-$T$ response since $\rho \to 0$ as $T \to 0$. We can say the effects of $J$ on
the response are negligible at high-$T$.
}

\subsubsection{Susceptibilities for $A_{1g}$ and $B_{1g}$ irreps}

Experimentally, it is possible to identify the irrep to which the order parameter belongs by applying a strain with a
particular irrep of strain and searching for a divergence in the temperature profile. In the case of uniaxial strain
along the x-axis the strain can be decomposed into the $A_{1g}$ and $B_{1g}$ irreps. In this section we examine
the strain-resistivity linear response function for the $A_{1g}$ and $B_{1g}$ irreps defined in terms of the hopping
strain as
\begin{eqnarray}
    \chi_{A_{1g}} &\equiv& -\Big (\frac{\rho'_{xx}+\rho'_{yy}- 2\rho_{xx}}{2\rho_{xx}} \Big ) 
                 \Big / \Big (\frac{\delta t}{t} \Big ) = \frac{\chi_{XX}+\chi_{YY}}{2} \;, \label{normalized-chiA1g} \nonumber \\
    \chi_{B_{1g}} &\equiv& -\Big (\frac{\rho'_{xx}-\rho'_{yy}}{\rho_{xx}} \Big ) 
                 \Big / \Big (\frac{\delta t}{t} \Big ) = \chi_{XX}-\chi_{YY} \;, \label{normalized-chiB1g}
\end{eqnarray}
respectively. 
%It is also useful to record the {unnormalized} susceptibility
%\begin{eqnarray}
%    \chi^{(u)}_{A_{1g}} \equiv -\Big ({\rho'_{xx}+\rho'_{yy}- 2\rho_{xx}} \Big ) 
%                 \Big / \Big (\frac{\delta t}{t} \Big ) \;. \label{unnormalized-chiA1g}
%\end{eqnarray}

\begin{figure}[h]
    \includegraphics[width=0.99\columnwidth]{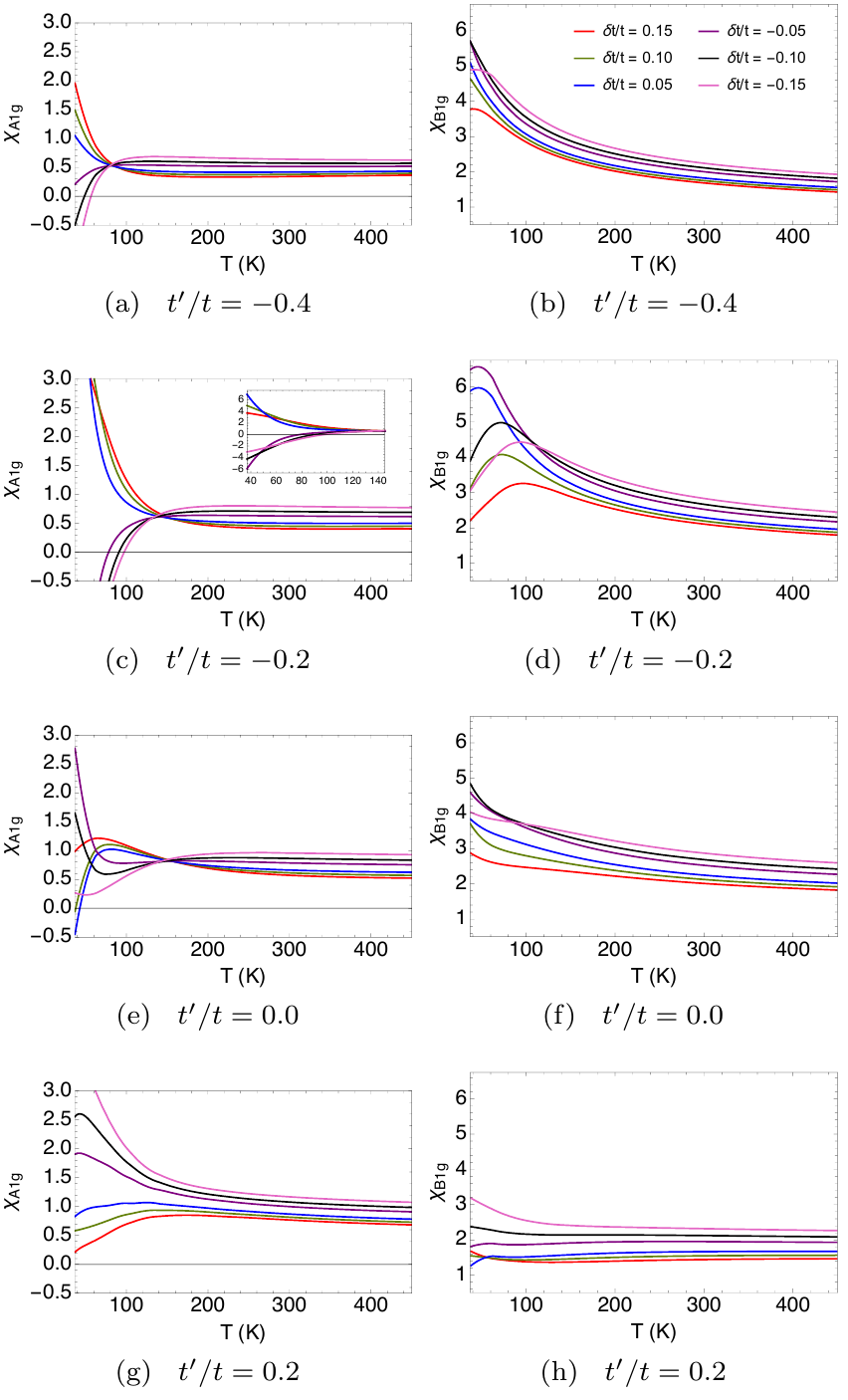}
    \caption{The normalized strain-resistivity susceptibilities from \disp{normalized-chiA1g} versus $T$ for the
        $A_{1g}$ and $B_{1g}$ irreducible representations at filling $n=0.85$ at various $t'$ and $\delta t / t$. All
        the figures share a legend. For various  $\delta t/t$,  and for  $T\gssim 100$ all the susceptibilities  approach  each other in  two sets, one for $\delta t/t>0$ and another slightly displaced set for  $\delta t/t<0$. They splay apart  at low T thus displaying strong non-linearity in the Fermi liquid regime.  }
    \label{fig:susrhoA1gnd085}
\end{figure}

In \cref{fig:susrhoA1gnd085} we present the normalized strain-resistivity response functions at
optimal density $n=0.85$ for various $t'$ and $\delta t / t$. In this picture the $A_{1g}$ and $B_{1g}$ irreps play the
roles of a center of mass coordinate and a relative coordinate, respectively. Together the two susceptibilities
characterize the shift of in-plane resistivity as a result of an arbitrary in-plane strain. 
%Note that in the unstrained system the $\rho_{xx} = \rho_{yy}$, and hence the $B_{1g}$ {irrep} vanishes for the unstressed system. 
Recall that since
the resistivity vanishes as $T \to 0$, the $A_{1g}$ and $B_{1g}$ susceptibilities are also enhanced at low-$T$.

Examining the $A_{1g}$ susceptibilities in \cref{fig:susrhoA1gnd085} {(a,c,e,g),} one important feature stands out, namely, that for $T \gssim
100$K the response function is positive for all $t'$ and strains $\delta t/t$. This indicates that increasing a tensile
(compressive) strain for $T\gssim 100$K enhances (suppresses) the average of the anisotropic resistivities.

%Quantitatively this result can be understood by examining the average of the anisotropic hopping parameters ratios $(t_d / t_x + t_d / t_y)/2$, which correlates with average size of the $T_{FL}$ scale which is important since it marks the cross-over from Fermi-liquid-like to non-Fermi-liquid-like behavior in the system.  

{ We also see that at $T \sim 100$K with hole
    doping, i.e. $t'\leq 0$, the normalized susceptibilities become independent of the strain, and hence the response is
    in the linear regime (signaled by the convergence of all strain curves). The non-linear response at lower $T$ is
    interesting and potentially observable in experiments with varying strain. On the other hand for electron doping,
    i.e.  $t'>0$, we see non-linear behavior even at high $T$. Its origin is the extended Fermi-liquid regime which has
a higher crossover temperature scale. Summarizing, we find that the early departure from Fermi liquid behavior into a
strange metallic behavior in the hole doping favors an apparent linear response above 100K due to a change in scale.
Conversely we expect to see non-linearity extending to much higher $T$'s in electron-doped systems.}

%\subsubsection{Strain-Resistivity susceptibilities with $B_{1g}$ irrep}
%In this section, we shall examine the response function for the $B_{1g}$ irrep defined as
%\begin{eqnarray}
%    \chi_{B_{1g}} \equiv -\Big (\frac{\rho'_{xx}-\rho'_{yy}}{\rho_{xx}} \Big ) 
%    \Big / \Big (\frac{\delta t}{t} \Big ) \;, \label{normalized-chiB1g} \\
%\end{eqnarray}

From \cref{fig:susrhoA1gnd085}, we observe that the $B_{1g}$ susceptibilities for $T < 100$K are strongly dependent on the value
of $t'$ of the system. We find in hole-like materials ($t' \lessim 0.0$) there is a strong enhancement (the details of
which depend on the $\delta t/t$) in the susceptibility at low-T. In contrast, this feature is absent in
electron-like materials ($t' > 0.0$) where there is weaker correlation, higher $T_{FL}$, and hence
stronger quasiparticles. 

Focusing on the strain dependence, we see that at high-T the susceptibilities are {relatively} insensitive to $t'$ and generally
increases as we vary from a compressive to a tensile strain. There is also asymmetry in rate of change of
susceptibilities between a compressive and tensile strain as $|\delta t/t|$ is varied, i.e., the response function
changes more rapidly for tensile than compressive strains. 
%\sout{We can quantify this asymmetry by relating it to the difference in the ratio of anisotropic hopping parameters $t_d / t_x - t_d / t_y$, where the difference in the ratios correlates with the difference in the slopes of the linear-in-T regimes} \Mike{\sout{which provides important information about the temperature scale upon which we expect to find Fermi-liquid-like and non-Fermi-liquid-like behavior}.} 
Therefore the degree of anisotropy is higher for tensile strain than compressive strains of equal magnitude. 

Also, the $B_{1g}$ curves under compressive strain ($\delta t/t>0$) are closer to each other than those under tensile
strain for electron-doped systems, yet this spacing difference is less obvious in the hole-doped case. It means that a
tensile response tends to show stronger non-linearity, especially in electron-doped systems.

\begin{figure}[h]
    \includegraphics[width=0.99\columnwidth]{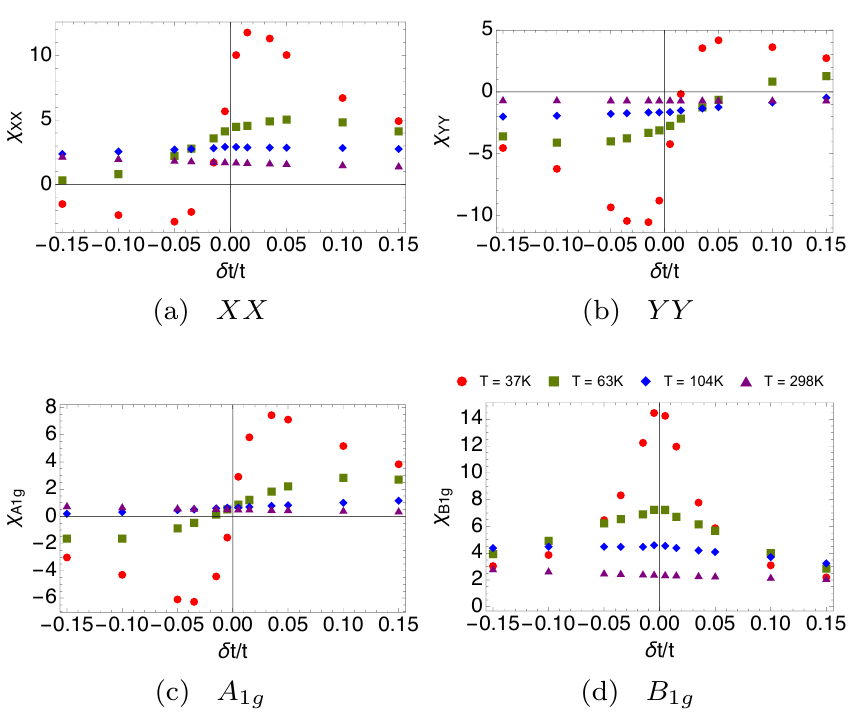}
    \caption{The strain-resistivity susceptibilities for various symmetries as a function of strain $\delta t/t$ at
    filling $n=0.85$ and $t'/t=-0.2$. (a) XX, longitudinal, \disp{norm-chiXX}; (b) YY, transverse, \disp{norm-chiYY};
    (c) $A_{1g}$ irrep, \disp{normalized-chiA1g}; (d) $B_{1g}$ irrep, \disp{normalized-chiB1g}.
     The susceptibilities are relatively strain independent above 100K, but show strong non-linearity at low T. It is noteworthy that for the lowest T shown, the  susceptibilities $\chi_{xx}, \chi_{yy}, \chi_{A_{1g}}$ change sign at or close to $\delta t/t=0$. 
      At higher $T$ this change of sign is lost.  The behavior of the nematic susceptibility $\chi_{nem}=\lim_{\epsilon_{xx}\to 0} (\alpha \chi_{xx})$ at low $T$ has thus the potential for a change of sign, depending on how we  choose a sufficiently small  $|\epsilon_{xx}|$ or   $|\delta t/t| $ for the purpose of taking the limit $\lim_{\epsilon_{xx}\to 0}$.
 }
    \label{fig:susVSdtrhond085}
\end{figure}

\subsubsection{Susceptibilities versus strain}

In \cref{fig:susVSdtrhond085}, we display the strain-resistivity response functions versus hopping strain for various
symmetries at $t'=-0.2t$ and $n=0.85$ (which is roughly the parameter set for LSCO cuprate material\cite{holedoped} at optimal density)
at four representative temperatures.
 
Here we approximate the variance in the linear response function as follows 
\begin{equation}\label{eq:chilinear}
\chi(T) = c_0(T)+ c_1(T)(\delta t/t)  + c_2 (T) (\delta t/t)^2 + \ldots\;. 
\end{equation}
In panel (a) and (b) we have longitudinal and transverse linear response functions, respectively, showing non-linear behavior
at low temperature which becomes more linear (as indicated by horizontal line) as the system warms. This non-linear behavior
at low-T can be understood as a result of the increasing importance of correlations as the system is cooled. Although
the longitudinal and transverse response functions differs considerably in magnitude, the curves are approximately
symmetric under inversion of the axes. In panels (a), (b) and (c) there is a wave-like oscillation which indicates the
presence of higher order terms, e.g., the $T=37$K curve in panel (a) appears to have $(\delta t/t)^3$ term competing with a linear
term.  Another interesting result we find that as the system cools the $B_{1g}$ response function appears diverge
at $\delta t/t = 0$ as $T\to 0$ suggests that any deviation from the point group symmetry of the square lattice produces a
finite resistivity response.

%\Mike{
%It is also interesting to comment upon linearity of linear response function. We observe that it is non-linear
%throughout the displayed temperature range. This observation can be understood as a artifact of strain-induced effects on the
%bare vertex $v_{\vec{k}}^{\alpha}$ in \disp{eq:sigmas} which also strongly influenced the longitudinal response.

 %In order to understand the non-linear behavior in longitudinal response is helpful to compare with the transverse
   % resistivity response function. Since the transverse response function is linear at high-T, we can surmise that the
    %non-linear behavior at high-T in the longitudinal response follows from the strain-induced changes to the band
    %structure. Additionally, the non-liner behavior of the longitudinal response function at $T \lessim 100$ K
    %can be attributed to affects of strain on both the correlation function and band structure.} 

%\sout{In panel (d) we observe that response function is nearly symmetric with respect to compressive and tensile strain; however, there is a slightly increased sensitivity to tensile strain.}

\subsection{Kinetic Energy for an x-axis strain}
In this section we explore the kinetic energy anisotropy induced by strain along the x-axis using ECFL theory. Since the
anisotropic kinetic energy can be related to measurements of the optical conductivity using the f-sum rule on the \tJ
model, this makes it an another interesting observable to explore. 

The total kinetic energy for a system under strain is computed as
\begin{equation} \label{eq:totkinetic}
    K_{\text{tot}} = \Big \langle \int^{\infty}_{-\infty}\rho_\G(\vec{k},\omega) \epsilon_{\vec{k}} d\omega \Big \rangle_{k}\;.
\end{equation}
This may be decomposed as follows:
\begin{equation}
    K_{\text{tot}} = K_{xx} + K_{yy} + K_{xy},
\end{equation}
where the cross kinetic energy $K_{xy}$ comes from the second neighbor interactions and is related to the dynamic Hall
conductivity. Additional information on the total kinetic energy can be found in the SM \cite{SM}. The longitudinal, transverse and cross kinetic energies are given by
\begin{align} \label{eq:anisoKE}
    K_{xx} &= \Big \langle \int^{\infty}_{-\infty} d\omega \rho_\G(\vec{k},\omega) \epsilon_{k_x} \Big \rangle_{k} \\
    K_{yy} &= \Big \langle \int^{\infty}_{-\infty}d\omega \rho_\G(\vec{k},\omega) \epsilon_{k_y} \Big \rangle_{k} \\
    K_{xy} &= \Big \langle \int^{\infty}_{-\infty} d\omega \rho_\G(\vec{k},\omega) \epsilon_{k_{xy}} \Big \rangle_{k}
\end{align}
where
\begin{align}
    \epsilon_{k_x} &= -2t_{x}\cos(k_x a) \\
    \epsilon_{k_y} &= -2t_{y}\cos(k_y b) \\
    \epsilon_{k_{xy}} &= -4t_d \cos(k_x a) \cos(k_y b)\;.
\end{align}
In the \tJ model the anisotropic kinetic energies $K_{\alpha}$, where $\alpha = xx$, $yy$ and $xy$, are related to the
optical conductivity $\sigma_{\alpha}$ by the following sum rule  
\begin{align}\label{eq:optweight}
    \Re e \int_{0}^{\infty} \sigma_{\alpha}(\omega) d\omega &= -K_{\alpha} e^2\;,
\end{align}
where $e$ is the electrical charge. $K_{\alpha} e^2$ sets the scale of the optical conductivity, i.e.,
\begin{align}
    -\frac{1}{K_{\alpha}e^2} \Re e \int_{0}^{\infty} \sigma_{\alpha}(\omega)d\omega = 1\;.
\end{align}
The optical conductivity in the DC limit $\sigma_{\alpha}(0)$ relates to the DC resistivity as follows: $\rho_\alpha(0)
= 1 / \sigma_\alpha(0)$. For the anisotropic kinetic energy, we calculate and quote the following objects:

\begin{itemize}
\item $K_{xx}'$ is the strained version of longitudinal kinetic energy. 
\item $K_{yy}'$ is the strained version of transverse kinetic energy. 
\item We call $K_{xx}$ without a prime the tetragonal result. It is the same as $K_{yy}$.
\item We present $A_{1g}$ : $$-\frac{K'_{xx} + K'_{yy} - 2K_{xx}}{2K_{xx} (\delta t / t)} \; \text{vs} \; \text{T} $$  
\item We present $B_{1g} : -(K'_{xx} - K'_{yy}) / (K_{xx} \delta t / t)$ vs T  
\end{itemize}

\subsubsection{Raw kinetic energies}

\begin{figure}[h]
    \includegraphics[width=0.99\columnwidth]{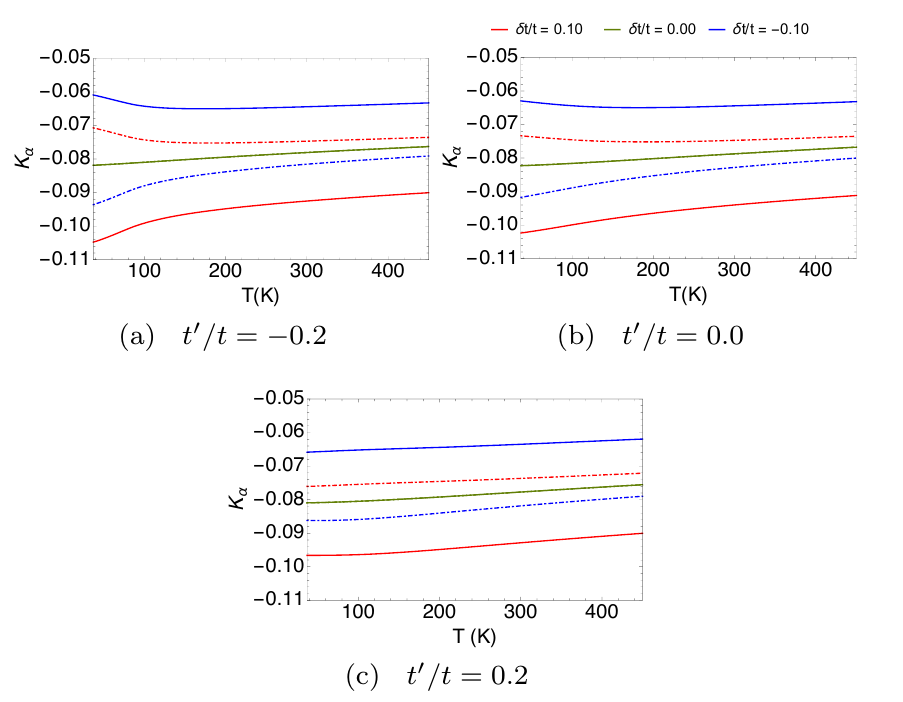}
    \caption{Anisotropic kinetic energies $K_{\alpha}$ versus $T$ for $K_{xx}$ (solid) and $K_{yy}$ (dashed) at filling $n=0.85$,
    for various $t'$ and at three representative strains: compressive strain, no strain, and tensile strain. Note that
$K_{xx} = K_{yy}$ in absence of strain. All figures share a legend.}
    \label{fig:KxxKyynd085}
\end{figure}

From Eq.~(\ref{eq:anisoKE}) we calculate the anisotropic kinetic energies $K_{\alpha}$ as a function of temperature at
optimal density for a representative range of cuprate materials $t'$ and hopping strains $\delta t / t$ as shown in
 \figdisp{fig:KxxKyynd085}. The main observation is that a compressive (tensile) strain suppresses (enhances) the longitudinal kinetic energy and vice versa for the transverse kinetic energy response
with a smaller magnitude of variation. The variation in the longitudinal kinetic energy can be understood as combination
of changes in the band structure parameter $t_x$ and correlations. On the other hand, the transverse kinetic energy is
dominated by changes to the correlation function since the parameter $t_y$ is unmodified by x-axis strain. There is
little $T$-dependence with exception to a slight broadening of the range of the response at low-$T$ as the
$T_{FL}$ is reduced. The $t'$-dependence is also weak because $K_{xx}$ and $K_{yy}$ does not explicitly depend on $t'$
but through the spectral function.

\subsubsection{Strain-kinetic-energy susceptibilities}

\begin{figure}[h]
    \includegraphics[width=0.99\columnwidth]{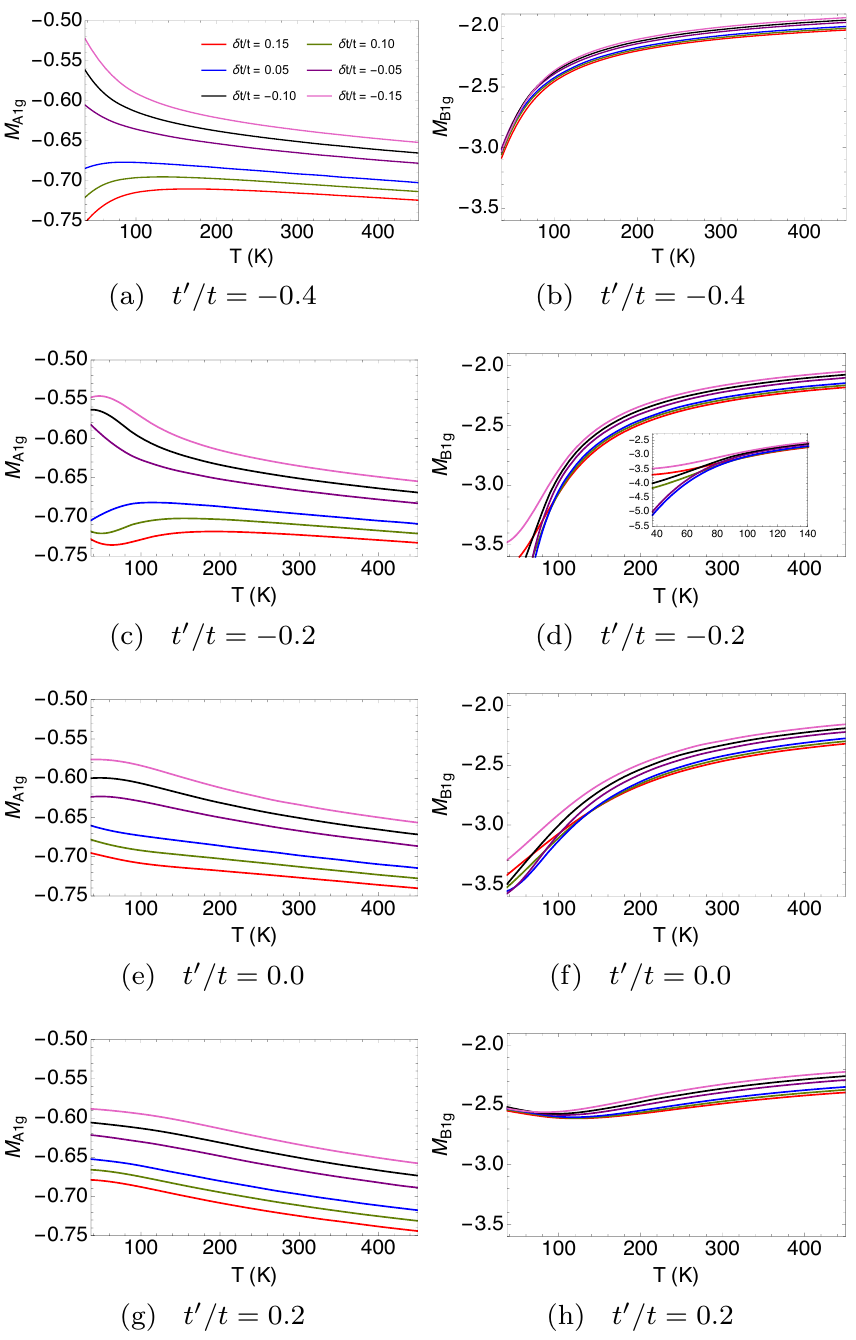}
    \caption{The normalized strain-kinetic-energy susceptibilities vs $T$ for the $A_{1g}$ and $B_{1g}$ irrep as defined in
    Eqs.~(\ref{normalized-MA1g}) and (\ref{normalized-MB1g}) at filling $n=0.85$, for various $t'$ and $\delta t / t$. All figures share a legend.}
    \label{fig:susKA1gnd085}
\end{figure}

In analogy with elastoresistance, we compute the so-called normalized strain-kinetic-energy response
function, which measures the change in kinetic energy with respect to a strain. We shall focus on the normalized
strain-kinetic-energy response functions for the $A_{1g}$ and $B_{1g}$ irrep since measurements of these symmetries are
sensitive to a break in the 4-fold rotation symmetry of a square lattice. Explicitly the response
functions are defined in terms of hopping strain as
\begin{eqnarray}
    M_{A_{1g}} &\equiv& -\Big (\frac{K'_{xx}+K'_{yy}-2K_{xx}}{2K_{xx}} \Big ) 
        \Big / \Big (\frac{\delta t}{t} \Big ) \;, \label{normalized-MA1g} \\
    M_{B_{1g}} &\equiv& -\Big (\frac{K'_{xx}-K'_{yy}}{K_{xx}} \Big ) 
        \Big / \Big (\frac{\delta t}{t} \Big ) \;, \label{normalized-MB1g}
\end{eqnarray}
where the sign is imposed so that susceptibility defined in terms of hopping strain matches its counterpart defined in terms of
conventional strain. \figdisp{fig:susKA1gnd085} displays the normalized strain-kinetic-energy susceptibilities as a
function of temperature for the $A_{1g}$ and $B_{1g}$ irrep at optimal density for various $t'$ and $\delta t / t$. 
%Here the temperature range begins below the typical $T_c$ of cuprate materials, forgetting the superconducting phase
%transition to show the normal state properties hidden beneath, and it warmed to above room temperature. 
%The $A_{1g}$ irrep susceptibility signals a change in the sum of anisotropic kinetic energies $K_{xx} + K_{yy}$ with respect to the
%sum of the strains $\epsilon_{xx} + \epsilon_{yy}$.  
%In the \tJ model the kinetic energy is negative and $K_{xx}=K_{yy}$ in the unstrained system. 
The $A_{1g}$ irrep susceptibility signals a change in the sum of anisotropic kinetic energies $K_{xx}$ + $K_{yy}$ with respect to the hopping change.
The $A_{1g}$ susceptibility shows that tuning the strain from tensile to compressive increases rather uniformly the magnitude of the anisotropic kinetic energy, i.e., strain enhances the overall optical weight
from \disp{eq:optweight}. Analysis of the longitudinal and transverse components are in the SM \cite{SM}.
%and that the response function varies non-linearly with respect to strain throughout the
%displayed temperature range with only a small variation in magnitude of the curves at $T<100$K.

%\sout{
%We observe that tensile strain uniformly reduces the total kinetic energy of the system and a compressive strain
%enhances it. Here, we will confine our attention to the normalized cases, Fig.~\ref{fig:susKA1gnd085}(e)-(h). We
%observe that the response functions is non-linear throughout the displayed temperature range for all $t'$. The even
%spacing between curves indicates that the function is non-linear to second order and the function monotonically increases
%in absolute intensity as the system is warmed and as the strain is increased form negative to positive.}

%In Fig.~\ref{fig:susKB1gnd085}, we display the normalized strain-kinetic-energy susceptibility versus $T$ for the
%$B_{1g}$ irrep, at optimal density ($n=0.85$) at various $t'$ and $\delta t /t$. 
The
$B_{1g}$ susceptibility is characterized as the difference in the kinetic energies $K_{xx} - K_{yy}$ with respect to the hopping change. Thus a non-zero value for the $B_{1g}$ irrep signals an anisotropy
between the two directions.  We observe that the response function for the $B_{1g}$ irrep is strongly $t'$ dependent.  For $t' =-0.4$, the response functions is
nearly linear at all temperatures. We point out a curious feature for $t'=-0.2$ curve where at high-$T$ the system is linear whereas at low-$T$ the system is non-linear, but it nearly symmetric with
respect to a compressive or tensile strain of similar magnitude. At high-$T$ for all $t'$ the system is monotonic with
respect to strain. For $t' \geq 0$ there is little variation in the response function across the temperature range and
it appears to become increasingly non-linear as the system is warmed due to the reduction in the scale of variation.

\subsubsection{Strain-kinetic-energy susceptibility versus strain}

\begin{figure}[ht]
    \includegraphics[width=.99\columnwidth]{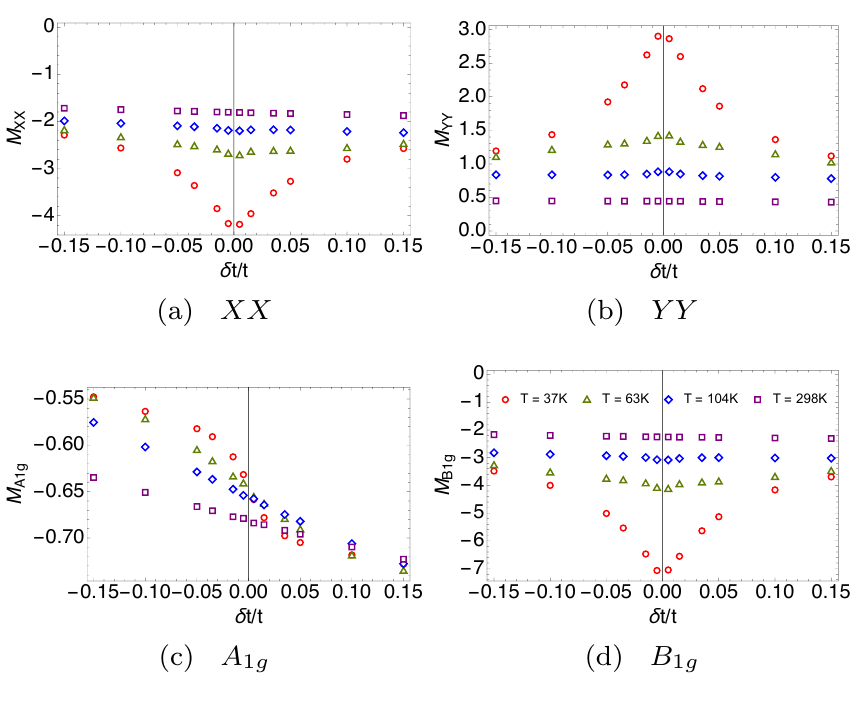}
    \caption{The strain-kinetic-energy susceptibilities versus $\delta t/t$ at filling $n=0.85$ and $t'/t=-0.2$ at four
        representative temperatures. All figures share a legend.  (a) $XX$, longitudinal, Eq.~(\ref{MXX}); (b) $YY$,
        transverse, Eq.~(\ref{MYY}); (c) $A_{1g}$ irrep, Eq.~(\ref{normalized-MA1g}); (d)
        $B_{1g}$ irrep, Eq.~(\ref{normalized-MB1g}).} 
    \label{fig:susVSdtKnd085}
\end{figure}

We now present strain-kinetic-energy susceptibility as a function of strain at optimal density
($n=0.85$) and $t'=-0.2t$ for $XX$, $YY$, $A_{1g}$, $B_{1g}$ symmetries at various $T$ (see
Fig.~\ref{fig:susVSdtKnd085}), where we define the longitudinal and transverse response functions as
\begin{eqnarray}
    M_{XX} \equiv -\Big (\frac{K'_{xx}-K_{xx}}{K_{xx}} \Big ) 
        \Big / \Big (\frac{\delta t}{t} \Big ) \;, \label{MXX} \\
    M_{YY} \equiv -\Big (\frac{K'_{yy}-K_{yy}}{K_{xx}} \Big ) 
        \Big / \Big (\frac{\delta t}{t} \Big ) \;. \label{MYY}
\end{eqnarray}
respectively. Like the resistivity case, $M_{A_{1g}}=0.5\times(M_{XX} + M_{YY} )$ and $M_{B_{1g}}=M_{XX} - M_{YY}$. 
%\sout{Here, we can examine the linearity of the response function
%at a fixed $T$. We note that a horizontal line shape is characteristic of a linear function, a sloped line is indicative
%of a second order term and a non-zero curvature signals higher order terms. For $A_{1g}$, we observe that the irrep is
%weakly non-linear up to room temperature, it grows increasingly non-linear as the system is cooled and the curve is
%anti-symmetric with respect to hopping strain. For the $B_{1g}$ irrep, we note that the response function is linear at
%room temperature; however, it becomes non-linear as the temperature is dropped, but in this case the response function
%is symmetric with respect to strain. The $A_{1g}$ irrep corresponds to the scale of the total optical conductivity,
%$\sigma_{xx}+\sigma_{yy}$, and it indicates that the total optical conductivity increases as we go from a tensile strain
%to a compressive strain. The $B_{1g}$ irrep signals the degree of anisotropy in the kinetic energies, i.e., a non-zero response
%indicates that the kinetic energy is more sensitive either in the longitudinal or transverse direction. The
%anti-symmetry of $A_{1g}$ together with the symmetry of $B_{1g}$ tell us that the sensitivity is symmetric for
%compressive and tensile strains.}

We find that at low temperatures, decreasing the magnitude of the strain increases the strength of the longitudinal response function in panel
    (a) and the response function is symmetric with respect to both strain types. The transverse response function in panel
    (b) shows a similar symmetry between tensile and compressive strains with a flipped sign.
     Therefore we find that a compressive strain for the $A_{1g}$ response function [panel (c)] depletes the in-plane optical weight and vice versa for
    a tensile strain. The $B_{1g}$ response function is similar to the longitudinal and transverse only more intensive
    and it signals an enhanced (suppressed) anisotropy between in-plane kinetic energies for a compressive (tensile)
    strains. In all cases the response function is approximately linear at room temperature (297K) and becomes increasingly non-linear as the system
    cools. In comparing panels
(b), (c), and (d) we see strong similarity between their respective responses. This is expected since strain merely
shifts kinetic energy versus temperatures curves up and down. Also, it appears to diverge for small strains as $T\rightarrow 0$.

\begin{figure}[h]
    \includegraphics[width=.99\columnwidth]{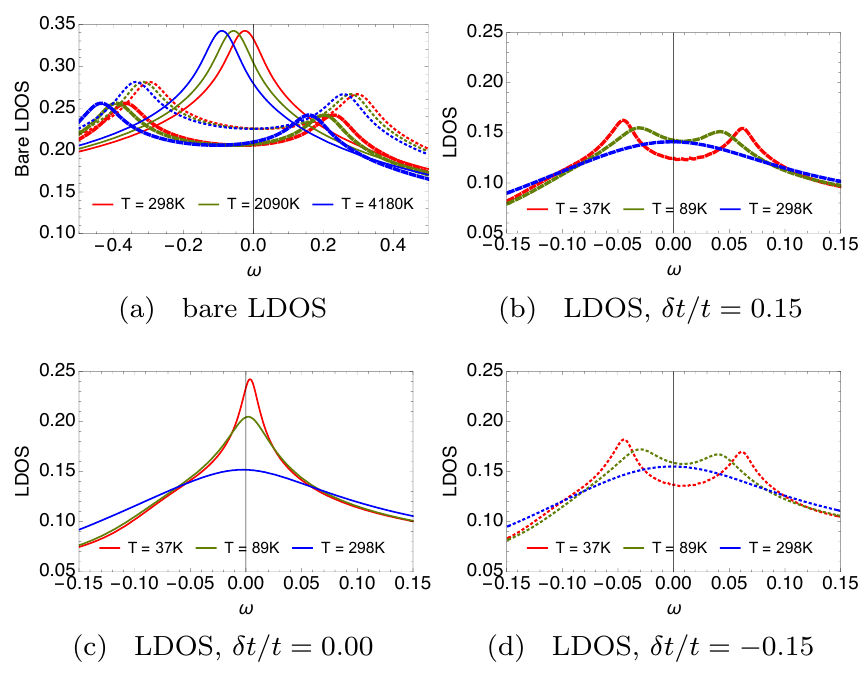}
    \caption{The local density of states for (a) the non-interacting {(band-structure)} and (b)-(d) interacting
        system {($t$-$J$ model)} at optimal filling
        ($n=0.85$), $t'=-0.2$t, for various temperatures and at three characteristic strains: $\delta t / t = 0.15, 0.00,
    -0.15$ (thick dashed, solid, thin dashed).}
    \label{fig:LDOSTvariation}
\end{figure}

\subsection{The local density of states for an x-axis strain}

The local density of states (LDOS) is also very interesting since it can be measured using STM probes. We present
results on how the LDOS changes with strain, and the related susceptibilities. We argue that if experiments are done on
resistivity variation as well as LDOS variation with strain, we can bypass the need for measuring strain accurately and
of estimating the parameter $\alpha$ in \disp{alpha}. The LDOS is calculated as $\rho_{G\text{loc}}(\omega) = \langle
\rho_{G}(\vec{k},\omega) \rangle_{k}$ where averaging over the Brillouin zone is implied, and $G \to g$ is the free
Green's function {(i.e., band structure)} which gives the bare LDOS and the ECFL Green's function $G \to \G$ gives
the LDOS for the {$t$-$t'$-$J$ model.}

In this section we calculate the normalized change in the local density of states and quote the following:

\begin{itemize}
\item {$\rho_{g\tx{loc}}'(\omega) = \langle \rho_{g}(\vec{k},\omega)\rangle_k$} is the bare LDOS for a strain along the x-axis 
\item {$\rho_{\G\tx{loc}}'(\omega) = \langle \rho_{\G}(\vec{k},\omega)\rangle_{k}$} is the interacting LDOS for an x-axis strain 
\item $\rho_{g\tx{loc}}$ without a prime refers to the tetragonal result and similarly for $\rho_{\G\tx{loc}}$.
\item We present $(\rho'_{g\tx{loc}} - \rho_{g\tx{loc}}) / (\rho_{g\tx{loc}} \delta t / t)$ vs $\omega$  
\item We present $(\rho'_{\G\tx{loc}}- \rho_{\G\tx{loc}}) / (\rho_{\G\tx{loc}}\delta t / t)$ vs $\omega$
\end{itemize}

\subsubsection{$T$ variation}
In Fig.~\ref{fig:LDOSTvariation}, we display the LDOS at optimal density ($n=0.85$) and $t'=-0.2$ for various
temperatures at three characteristic strains: a compressive strain (thick dashed), unstrained (solid) and tensile strain
(thin dashed). We compare the LDOS for a non-interacting system [panel (a)] to a system with electron-electron interaction
[panels (b)-(d)]. We find over large temperature scales that curves for the bare LDOS shifts to left along the $\omega$-spectrum upon warming, leaving the line shape intact. In
contrast with the bare LDOS, we see that warming the LDOS for the interacting system in panel (c) completely smooths and broadens the
LDOS peaks for all strains, and slightly shifting them left. This is consistent with previous findings that interactions significantly lower the Fermi liquid temperature $T_{FL}$\cite{SP}. 
We note that strain inverts the LDOS peak at low-T, leaving behind a
pair of cusps at a reduced height. This is an artifact of the anisotropy of hopping parameters since it also shows up in the bare case.
{
\subsubsection{$J$ variation}

\begin{figure}[h]
    \includegraphics[width=.99\columnwidth]{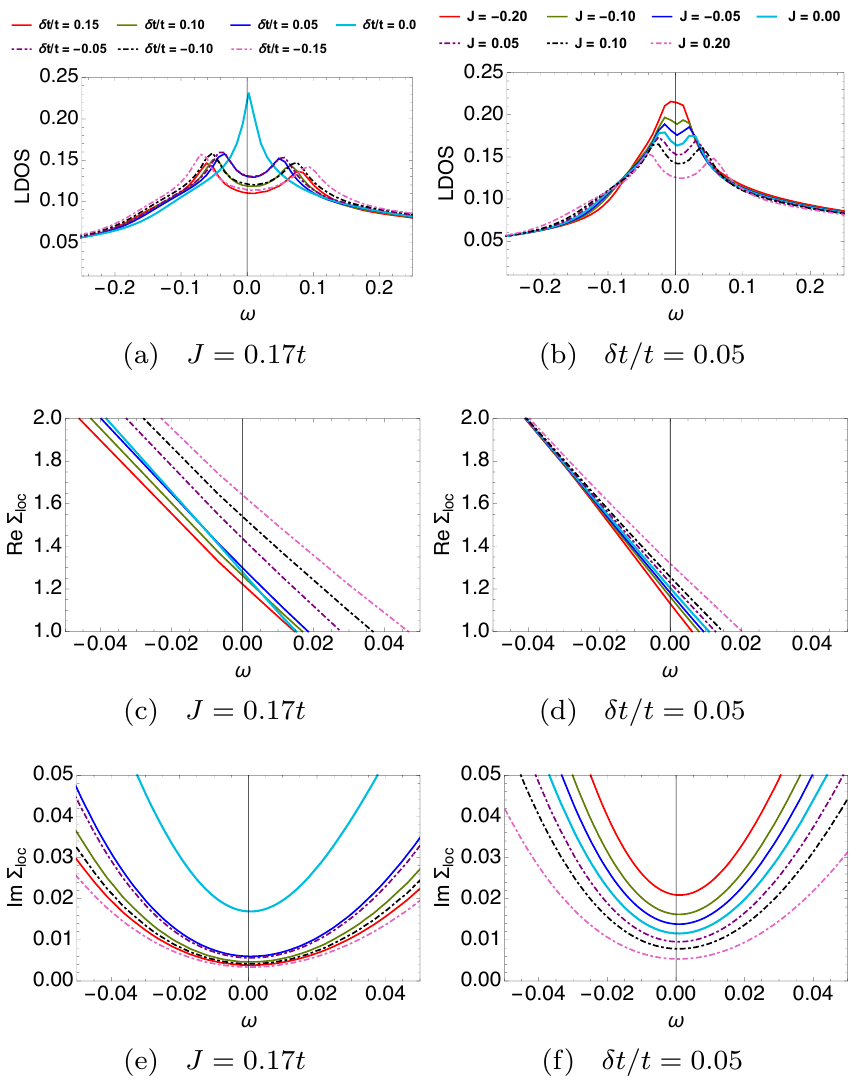}
    \caption{ (a,b) The LDOS; (b,c) real part of local Dyson self-energy; (e,f) the imaginary part of local Dyson self-energy
    for parameter set $n=0.85$, $T=37$K, $t'=-0.2t$ with varying $\delta t / t$ (LHS) and varying $J$ (RHS). Figures
    (a,c,e) and (b,d,f) share a legend, respectively.}
    \label{fig:LDOSJvariation}
\end{figure}

In \figdisp{fig:LDOSJvariation}, we turn on the exchange parameter $J$ and examine the LDOS. We also find it useful to
examine the self-energy of the system. We define the Dyson self-energy $\Sigma$ as
\begin{equation}
    \G(k) = \frac{1}{\omega + \chem -\epsilon_{\vec{k}} - \Sigma(k)}\;
\end{equation}
Here we use the shorthand $\Sigma = \Sigma'+i\Sigma''$ to denote the real and imaginary parts of a complex function. In
terms of the spectral function, self-energy imaginary part is  
\begin{equation}
    \Sigma''(k) = \frac{-\pi\rho_{\G}(k)}{[\G'(k)]^2 + [\pi\rho_{\G}(k)]^2}\;,
\end{equation}
where $\Re e \G = \G'$ is found by taking the Hilbert transform of $\Im m \G = \G''$ and we can find $\Sigma'$ in the
same manner. In Figure~\ref{fig:LDOSJvariation} (c-f) we display the Dyson self-energy averaged over the Brillouin zone $\Sigma_{\text{loc}}(\omega) = \langle \Sigma
(\vec{k},\omega )\rangle_{k} $.

Turning on the exchange parameter in  \figdisp{fig:LDOSJvariation} (a) has a small, but visible
effect on LDOS at low-$\omega$ when compared to S-Fig.~6(b) of the SM \cite{SM}. For panel (c) we see that varying strain from compressive ($\delta t/t > 0$) to tensile ($\delta
t/t < 0$) shifts the average quasi-particle states to higher energies and panel (e) shows that increasing the intensity
of the strain produces quasi-particles with higher and sharper peaks. In panels (b,d,f) we see that varying $J$ from
ferromagnetic (negative) to anti-ferromagnetic (positive) splits a single LDOS peak into two, shifts the average
quasi-particle states to higher energies, and narrows the quasi-particle peaks.}

\subsubsection{$t'$ variation}

\begin{figure}[h]
    \includegraphics[width=.99\columnwidth]{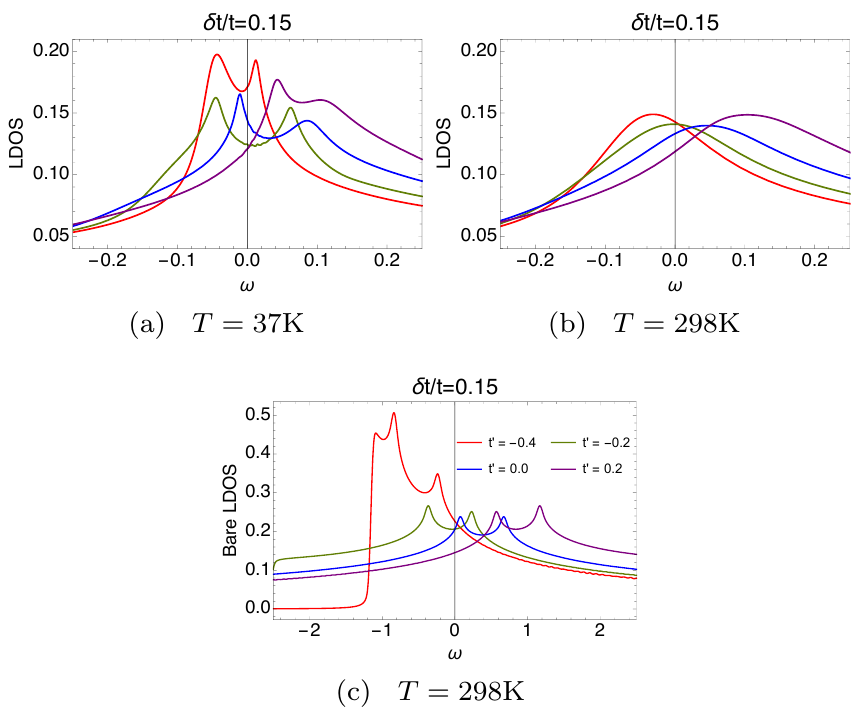}
    \caption{The local density of states versus frequency at optimal filling ($n=0.85$), for a compressive strain
        ($\delta t/t = 0.15$) at various $t'$. (a)-(b) The interacting system {(\tJ model)} at $T=37$K and $T=298$K, respectively. (c) The
    non-interacting {(band-structure)} system at $T=298$K. All figures share the same legend.}
    \label{fig:LDOStpvariation}
\end{figure}

In Fig.~\ref{fig:LDOStpvariation}, we examine the LDOS from a different vantage point by looking at the $t'$ dependence
for a system at optimal density ($n=0.85$), for a compressive strain of $\delta t / t = 0.15$, at various $t'/t$. In
panel (c), we show the bare LDOS at room temperature as a reference for the interacting system. In panels (a) and (b),
we display interacting system at $T=37$K and $T=298$K, respectively. Upon inspection it appears the primary role that
$t'$ plays is to shift the energy band along the spectrum. As previously noted, warming the interacting system to room
temperature smooths and broadens the characteristic LDOS peaks for all strain types and at all $t'$ while leaving their
position in the spectrum fixed. Even though the relative position of different $t'$ curves remain unchanged as the
interactions are turned on, we note that strong correlations renormalizes the bare band into a smaller energy region.
Comparing panels (a) and (b) fixed at $t'=-0.4, -0.2$, we observe that LDOS peak height is more strongly suppressed at a
lower $t'$. This is consistent with previous studies\cite{PS} on the unstrained interacting system, and it indicates
that a smaller $t'$ has a lower Fermi-liquid temperature scale and hence it is less robust to heating. For further
analysis of the strain dependence see the SM \cite{SM}.

\begin{figure}[h]
    \includegraphics[width=.99\columnwidth]{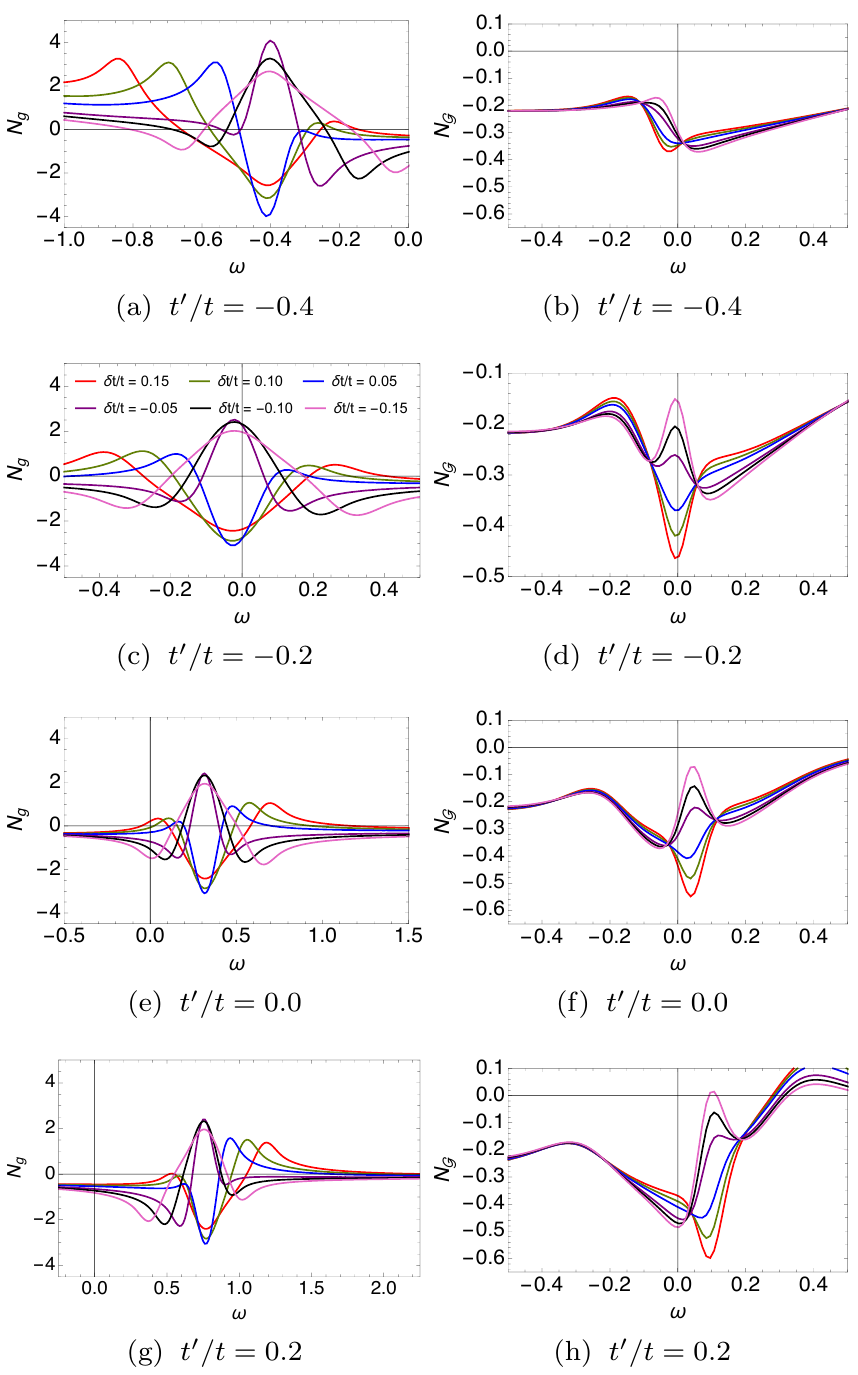}
    \caption{The LDOS susceptibility versus frequency at optimal filling $n=0.85$, at room temperature ($T=297K$), for various
        $t'$ and $\delta t / t$. The non-interacting {(band-structure)} system (LHS) and interacting system
        {(\tJ model)}
    (RHS) from \disp{susbareLDOS} and \disp{susLDOS}, respectively. All the figures share a legend.}
    \label{fig:susLDOSnd085}
\end{figure}

\subsubsection{Susceptibilities}

Next, we examine the normalized response function of LDOS of the non-interacting and
interacting system, respectively, defined as
\begin{eqnarray}
    N_{g} \equiv \Big ( \frac{\rho'_{g\tx{loc}}-\rho_{g\tx{loc}}}{\rho_{g\tx{loc}}} \Big ) 
    \Big / \Big ( \frac{\delta t }{t} \Big ) \;, \label{susbareLDOS} \\
    N_{\G} \equiv \Big ( \frac{\rho'_{\G\tx{loc}}-\rho_{\G\tx{loc}}}{\rho_{\G\tx{loc}}} \Big ) 
    \Big / \Big ( \frac{\delta t}{t} \Big ) \;. \label{susLDOS}
\end{eqnarray}

In Fig.~\ref{fig:susLDOSnd085}, we plot the LDOS susceptibility for a non-interacting and interacting system at room
temperature at optimal density for various $t'$. We observe that the response function is linear at all frequencies
except near the LDOS peak and, although not shown in the figure, at the band edges. Regardless of the presence of
interaction, we note that the susceptibility is enhanced by tensile strain near the LDOS peak and reduced by a
compressive strain.

\subsubsection{Susceptibility versus strain}

\begin{figure}[h]
    \includegraphics[width=.99\columnwidth]{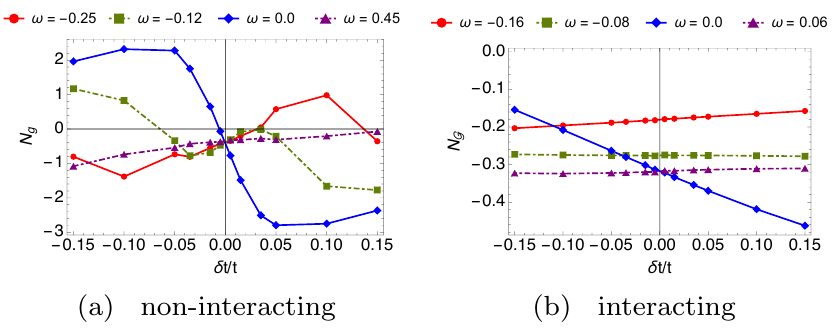}
    \caption{The LDOS susceptibility versus strain at optimal filling $n=0.85$, at room temperature ($T=298K$), for
    $t'=-0.2t$, at a few representative frequencies $\omega$ in units of $t$. (a) The non-interacting system
    {(band-structure)} in Eq.~(\ref{susbareLDOS}). (b) The interacting system {(\tJ model)} in Eq.~(\ref{susLDOS}).}
    \label{fig:susLDOSVSdt}
\end{figure}

Changing up the perspective, we explore the LDOS susceptibility now as a function of strain, at four representative
frequencies as seen in Fig.~\ref{fig:susLDOSVSdt}. 
%\sout{A curve that is constant is characteristic of a linear response function. We see that at $\omega = 0$ the response function is non-linear to second order.} 
We can approximate
the variance in the linear response function in Eqs.~(\ref{susbareLDOS}) and (\ref{susLDOS}) as
\begin{equation}
    N(T) = c_0 (T) + c_1(T) (\delta t/t) + c_2(T) (\delta t / t)^2 + \ldots
\end{equation}
where $c_0$ is the linear term, $c_1$ is the second order term, and $c_2$ is the third order term of the
response. We see
that for the bare LDOS, Fig.~\ref{fig:susLDOSVSdt}(a), at $\omega = 0.45$ the system is nearly linear with $c_0 \approx
-0.5$ and $c_1 \approx 3$. The other presented frequencies appear to be non-linear with significant second and third
order terms. The LDOS susceptibility for interacting system [panel (b)] appears to be nearly linear everywhere except at
the location of LDOS peak ($\omega = 0$) which has a strong quadratic response, suggesting that at temperatures relevant to
experiments non-linear behavior is only observable at energies near the Fermi surface. Note that the second order scheme used here
is good for low energies but somewhat less reliable at high energies, $| \omega | \gssim k_B T_{FL}$.

\section{Summary and  Comments }
\label{sec:Conclusion}

\subsection{Summary }
In this work, we have applied the ECFL theory to study the effect of small strain on the
resistivity, kinetic energy, LDOS {and} their associated susceptibilities
 in the $t$-$t'$-$J$ model  \disp{Ham-0}  with various $t'$ at
$n=0.85$. These results are expected to be relevant to cuprate superconductors, especially single layered materials, where the calculated unstrained resistivities are  in good accord with {the experimental} data \cite{S-M-New}.
   
   Based on comparisons carried out earlier, the second order scheme of ECFL used here is expected to be reasonable in
   the density range $0.85\gssim n \gssim 0.80$ spanning an experimentally accessible range in cuprates. With improvements
   in the theoretical scheme, we expect that while resistivities themselves might not change too much, the related
   susceptibilities [involving division by the small resistivity as in \disp{nematic-chi}] could be more sensitive.

Our results exhibit in considerable detail the theoretically expected strain dependence of resistivity and LDOS as well as optical weight. The derived susceptibilities depend sensitively on the magnitude and sign of $t'$. Our results in Fig.~(2) and Fig.~(3) illustrate the quantitative change of the strain dependence due to varying the magnitude and sign of t'. We should stress that the {absolute scale} of $t$ is important in determining the T dependence. For illustration we have used $t=0.45$eV in the present paper while the more fine-tuned estimates in \cite{S-M-New} suggest a material dependent and somewhat larger value of $t$$\sim$$1$eV in most cases.

Our results can be converted to actual strains as in  \disp{nematic-chi-2}, with $\alpha$ in the range $\alpha \in \{2,5\}$. If data is available  one may ideally   eliminating $\alpha$ by measuring the strain dependence of the LDOS or the optical conductivity sum-rule.

%We find that the longitudinal resistivity gets higher under a compressive ($\delta t/t>0$) strain and lower under a tensile ($\delta t/t<0$) one for a $t'$, while the sign of change in the transverse resistivity depends on $t'$. In the exceptional case $t'/t=0.2$, the transverse resistivity is unchanged under small strain. 

\subsection{Comments on experiments}

The results found in \figdisp{fig:susrhoxxrhoyynd085} yield a magnitude of the nematic susceptibility $\chi_{nem}\sim
(1-5) \alpha$ for cuprates. Using the expected range of $\alpha\in\{2,5\}$, we find  $\chi_{nem}$$\sim$ 2-25). On the
other hand, iron based pnictide superconductors appear to have a considerably larger value for $\chi_{nem}$, e.g., in
Fig.~(3) of \refdisp{Fisher2012} the range $|\chi_{nem}|\lessim 650$ is reported, thus an order of magnitude greater
than our theoretical estimate for cuprates. While fluctuations may drive the magnitude of nematicity further upwards,
especially at some densities and temperatures, it appears that the baseline magnitude of this object is itself much
larger than expected in cuprates. For example in the four featureless curves of Fig.~(3) of \refdisp{Fisher2012} we
see that $|\chi_{nem}|\sim$200.

This magnitude indicates that the downfolding of the many bands of the pnictides to an effective single (or few) band
model must yield hopping parameters that are much more sensitive to strain than in cuprates. The different type of
quantum overlap of relevant atomic orbitals from those in cuprates are presumably the origin of this difference.  We
also note that the sharp peaks in $|\chi_{nem}|$ on varying T, as reported in \cite{Fisher2012,Fisher2016} are missing
in our results. Instead we have a monotonic increase of $|\chi_{nem}|$ and related susceptibilities as we cool the
system, as seen in \figdisp{fig:susrhoxxrhoyynd085} and \figdisp{fig:susrhoA1gnd085}. This increase is largely due to
the decrease of the (unstrained) resistivity with lowering $T$ in the Fermi liquid regime.

The sign of $\chi_{nem}$ presents a more subtle problem. In iron pnictides it is known to be sensitive to effective mass
anisotropy. In fact it changes sign with doping in certain hole-doped iron pnictides \cite{Blomberg}. Our single band
model lacks such an anisotropy and is therefore not appropriate to describe the elastoresistivity of iron pnictide
materials.

After initial submission of our manuscript, we came across the recent measurement of the elastoresistivity nematic
susceptibility in \refdisp{NewER} on the two layer cuprate Bi$2212$. In this experiment, the magnitude of the nematic
susceptibility is found to be in the range $|\chi_{nem}|\in \{ 2.5, 5\}$. This range is consistent with our theoretical
estimate. It is also smaller than the nematic susceptibility in iron pnictides by about two orders of magnitude.

The sign of the nematic susceptibility $\chi_{nem}$ (\disp{nematic-chi}) reported in \refdisp{NewER} implies  that the
resistivity {\it increases} in the direction of compression. This result has the opposite sign to our theoretical result
as seen in \figdisp{fig:susrhoxxrhoyynd085}. There we see that the theoretical resistivity {\it decreases} in the
direction of compression, although it does {\it increase} in the transverse direction.
It is possible that the two layer nature of Bi$2212$ might be responsible for this opposite sign. 
Also as noted in \figdisp{fig:susVSdtrhond085},
the behavior of the nematic susceptibility $\chi_{nem}=\lim_{\epsilon_{xx}\to 0} (\alpha \chi_{xx})$ at sufficiently low $T$ has  the potential for a change of sign, depending on how we  choose a sufficiently small  $|\epsilon_{xx}|$ or   $|\delta t/t| $ for the purpose of taking the limit $\lim_{\epsilon_{xx}\to 0}$. On the experimental side, a  more detailed T variation and examining the various susceptibilities listed in \figdisp{fig:susVSdtrhond085} should yield a more complete picture.

{The results
    found here should also motivate further studies of the strain variation of the 3-dimensional electronic bands of
    cuprates, towards computing strain variation of the resulting 2-dimensional bands found from projecting to a \tJ
    model. These would test the simple assumptions made here between strain and hopping parameters of a reduced
    2-dimensional model as presented in Eqs.~(\ref{t-versus-R-2}), (\ref{eq:hopparams}), (\ref{delta-t-epsilon}), and
    (\ref{signs}). {It} is {also} possible that under certain situations, the sign of $\alpha$ can even be changed, as a naive
    interpretation of the experiments of Ref.~\cite{NewER} suggests.}

We believe that it is important to study more extensive set of samples including single layer cuprates at various
compositions in future. It would also be useful to study the variations of resistivity along different axes, parallel
{\it and} transverse to the strain axis and  extend the studies to  various T's. This type of measurements would enable the construction of the symmetry adapted
susceptibilities as in  \figdisp{fig:susVSdtrhond085},  which provide a greater insight in to the results. It would also be of considerable interest  to measure
the variations of the LDOS and optical weight with strain, as emphasized above.
\\ \\

\section{\bf Acknowledgement:} 
We thank Professors  I. R.  Fisher and S. A. Kivelson for stimulating our interest in  this problem, and for helpful discussions.  We  thank Professor T. Shibauchi for helpful explanation  of  features of \refdisp{NewER}.
The
work at UCSC was supported by the US Department of Energy (DOE), Office of Science, Basic Energy Sciences (BES), under
Award No. DE-FG02-06ER46319. The computation was done on the comet in XSEDE \cite{xsede} (TG-DMR170044) supported by
National Science Foundation grant number ACI-1053575.

%\appendices
%\section{Summary}

%\bibliographystyle{apsrev4-1}
%\bibliography{References}

\begin{thebibliography}{99}
\bibitem{Http} {In the t-t'-J model the first and last term of Eq.~\ref{Ham-0}, referred to as the hopping
        $H_t$ and exchange $H_J$ term respectively, explicitly
        expanded as sum over neighbors is
    $$ 
    H_{t} +\chem \hat{N} = -t \sum_{<ij> \sigma} \widetilde{C}^{\dagger}_{i\sigma}
              \widetilde{C}_{j\sigma} - t'\sum_{\ll ij \gg \sigma}
              \widetilde{C}^{\dagger}_{i\sigma}\widetilde{C}_{j\sigma}\;,
    $$
    $$ H_J = \frac{J}{2}\sum_{<i j>} \left( \vec{S}_i.\vec{S}_j - \frac{1}{4} n_i n_j \right)\;,  
    $$
    where $<\!\!ij\!\!>$ is the sum over nearest neighbors, and $\ll \!\! ij \!\! \gg$ is the sum over second nearest
    neighbors. In terms of the Hubbard operators --- see Eq.~(\ref{eq:Hamiltonian})---this becomes 
    $$ 
    H_{t} + \chem \sum_{i\sigma} X^{\sigma\sigma}_{i} = -t \sum_{<ij> \sigma}
                X^{\sigma 0}_{i} X_{j}^{0\sigma} - t'\sum_{\ll ij \gg \sigma} {X}^{\sigma 0}_{i}X_{j}^{0\sigma}\;,
            $$
        \begin{align*}
            H_J &= \frac{J}{2}\sum_{<ij>\sigma}X^{\sigma\sigma}_i \nn \\
          &\;\;+ \frac{J}{4}\sum_{<ij>\sigma_1\sigma_2}
          \{X^{\sigma_1\sigma_2}_i X^{\sigma_2\sigma_1}_j 
          - X^{\sigma_1\sigma_1}_i X^{\sigma_2\sigma_2}_j \}\;. \nn
        \end{align*}
}
\bibitem{comment-1}  This can be seen e.g. \disp{eq:selfenergy2}, where the ``interaction'' term in the self-energy is
determined by the $\varepsilon_{\vec{k}}$'s, the Fourier transform of the hopping matrix elements.
\bibitem{HJ}{Here $J_{\vec{q}}$ is the Fourier transform of $J_{ij}$ over nearest neighbors. For a homogeneous and
translationally invariant system, the Fourier transform gives $J_{\vec{q}} = 2J_{x} \cos(q_{x}a) + 2J_y \cos(q_{y}b)\;$
on a rectangular lattice where $J_x$ and $J_y$ are the exchange parameter along the principle axes.}
\bibitem{materials} {We refer to $n=0.85$ as the optimal density since in a typical single-layer cuprate superconductors
    such as La$_{2-x}$SrCu O$_4$ (LSCO) --- to which the 2D $t$-$J$ model is most relevant the system --- obtains its
    maximum critical temperature $T_c$ at hole doping $\delta = 0.15$. Similarly, Ba$_2$Sr$_2$CaCu$_2$O$_{8+\delta}$
    (Bi2212) a double-layer curprate superconductor obtains its max $T_c$ at doping $\delta = 0.15$.}
\bibitem{Distance} {The unstrained system assumes a body-centered orthorhombic unit cell
$(a, b, c)$ where $a$, $b$ are the lattice constants of the base and $c$ is that of the height. In the expression for
resistivity $\rho_0$, $\fc_0$ corresponds to the interlayer separation $\fc_0 = c/2$ between copper-oxide planes.}
\bibitem{holedoped} {In high $T_c$ systems estimates give $t' \lesssim -0.27t$ Bi2212 and LSCO $t' \sim 0.16t$. We take
    $t'=-0.2t$ as a comprise between the two extremes.}
\end{thebibliography}

\twocolumngrid
\clearpage
%\newpage
%\begin{center}
%\textbf{\large Supplemental Materials: Title for main text}
%\end{center}
\onecolumngrid
\begin{center}
\textbf{ \large Supplementary Material:\\ 
    Theory of anisotropic elastoresistivity of \\
    two-dimensional extremely strongly correlated metals} \\ [.4cm]
     Michael Arciniaga$^1$, Peizhi Mai$^{2,1}$, B Sriram Shastry$^1$ \\
    \textit{\small $^1$Physics Department, University of California,  Santa Cruz, CA 95064, USA and \\} 
    \textit{\small $^2$Center for Nanophase Materials Sciences, Oak Ridge National Laboratory, Oak Ridge, TN, 37831-6494,
    USA \\}
    \text{\small (Dated:\;\today)} \\ [1cm]
\end{center}
\twocolumngrid

%%%%%%%%%% Merge with supplemental materials %%%%%%%%%%
%%%%%%%%%% Prefix a "S" to all equations, figures, tables and reset the counter %%%%%%%%%%
\setcounter{section}{0}
\setcounter{equation}{0}
\setcounter{figure}{0}
\setcounter{table}{0}
\setcounter{page}{1}
\makeatletter
\renewcommand{\theequation}{S\arabic{equation}}
\renewcommand{\bibnumfmt}[1]{[S#1]}
\renewcommand{\citenumfont}[1]{S#1}

\renewcommand{\figurename}{S-FIG.}
\renewcommand{\disp}[1]{Eq.~(\ref{#1})}
\renewcommand{\dispop}[1]{~{\bf (\ref{#1})}}
\renewcommand{\refdisp}[1]{Ref.~[S\onlinecite{#1}]}
\renewcommand{\figdisp}[1]{S-Fig.~\ref{#1}}

\section{Overview}
The Supplementary Material (SM) sections are organized as follows: In Sec.~\ref{S-sec:Resistivity} we expand the resistivity
analysis by discussing the role of different terms on the resistivity calculation under an x-axis strain. In
Sec.~\ref{S-sec:KineticEnergy} we examine the effects of strain on the total kinetic energy and
the longitudinal and transverse susceptibilities over a broad temperature range. In Sec.~\ref{S-sec:LDOS} we present
{supplemental figures} showing the effects of strain on the local density of states (LDOS) in different materials for
both non-interacting (bare band structure) and interacting (\tJ model) systems.

\section{Resistivity for an x-axis strain {: role of different factors}}
\label{S-sec:Resistivity}

Here we further explore the effects of strain on electrical resistivity and their associated susceptibilities in response
to an x-axis strain for an extremely correlated Fermi liquid \cite{S-ECFL,S-ECFL2} (ECFL). In order to gauge the relative
importance of strain in the resistivity calculation \cite{S-SP}, it is useful to study the effect of strain on each term in
conductivity calculation:

\begin{figure}[ht]
    \includegraphics[width=.99\columnwidth]{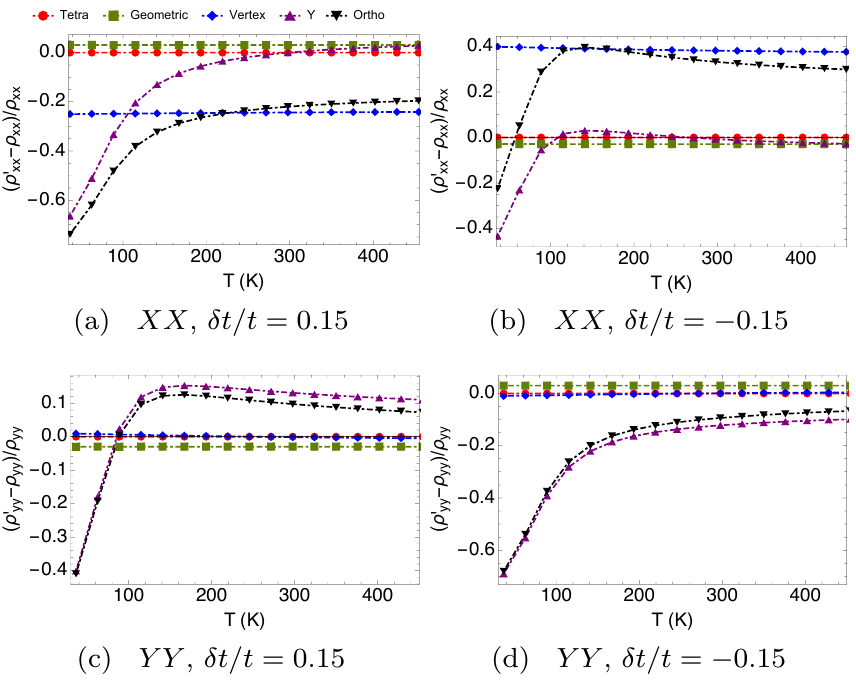}

    \caption{The normalized resistivity response vs $T$ isolating the effects of different terms in
    Eqs.~(\ref{S-eq:modsigmas1}) and (\ref{S-eq:modsigmas2}) for a holed doped material. The curve labeled
    ``Geometric'' gauges the sensitivity of resistivity to a direct change in lattice parameters $a$ and $b$.
    ``Vertex'' examines the effects of strain on $\partial \epsilon_{\vec{k}}/\partial k$ which is modified to be
    only a function of $t,t'$. ``$\Upsilon$'' isolates the effects of strain on $\rho_\G$. ``Tetra'' is the
    unstrained resistivity while ``Ortho'' includes the strain-induced effects from all the terms. All figures share a
    legend. The parameters set is filling $n=0.85$, $t'=-0.2t$ and strain of magnitude $|\delta t/t| = 0.15$.} 
    \label{S-fig:CasesHoleDoped} 
\end{figure}

\begin{align}    
 \sigma_{xx} &=  \Big \langle \Upsilon_{\vec{k}} \bigg (\frac{\partial \epsilon_{\vec{k} } }{\partial k_{1}} \bigg )^2 (a / b) \Big \rangle_k \;, \label{S-eq:modsigmas1} \\
 \sigma_{yy} &= \Big \langle \Upsilon_{ \vec{k} } \bigg (\frac{ \partial \epsilon_{\vec{k}} }{\partial k_{2} } \bigg )^2 (b / a) \Big \rangle_k \;. \label{S-eq:modsigmas2}
\end{align}
Furthermore we note that strain affects conductivity in two fashions: it produces a change (1) in the lattice constant
\begin{equation}
a \to a_0 + \delta a; \;\; b\to a_0; \;\; \rho \to \sqrt{2} a_0 + \frac{ \delta a}{\sqrt{2}}\;, \label{S-eq:lattice-constants}
\end{equation}
and (2) in the hopping parameters
\begin{align} \label{S-eq:hopparams}
t_{x} = (1 -  \alpha \; \epsilon_{xx}) \; t;  \; \;
t_{y} = t; \;\;
t_{d} = \Big (1 - \alpha \; \frac{\epsilon_{xx}}{2} \Big ) \; t'\;. 
\end{align}
The distorted lattice constants play the role of a dimensionless coefficient in Eqs.~(\ref{S-eq:modsigmas1}) and
(\ref{S-eq:modsigmas2}). The modified hopping parameters comes into play through the bare vertices and the spectral
function in $\Upsilon_{\vec{k}}$ (see Eq.~34 of the main text). Specifically, we compute the conductivity while
isolating the effects of strain to the following terms: (1) the spectral functions in $\Upsilon_{\vec{k}}$, (2) the
``vertex'' $\partial \epsilon_{\vec{k}}/\partial k_{i}$ for $i = 1 , 2$, and (3) the distorted lattice constants $a,b$,
denoted ``Geometric'' which separates out the strain-induced effects on the correlations, band structure and geometry,
respectively, and we record the sensitivity of conductivity to each term.

\begin{figure}[h]
   \includegraphics[width=.99\columnwidth]{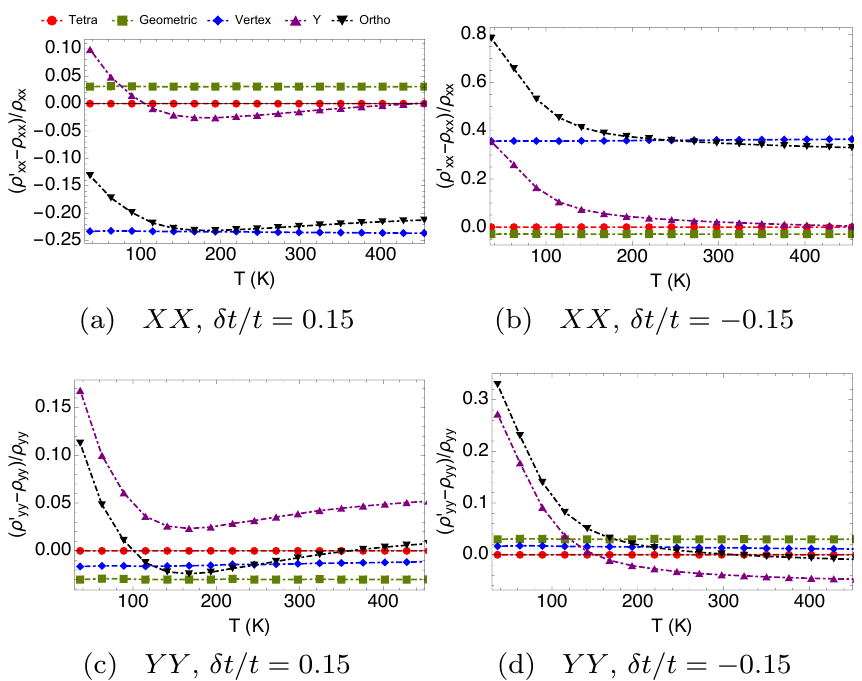}
    \caption{The effects of strain on the normalized resistivity response vs $T$ isolating different terms in
    Eqs.~(\ref{S-eq:modsigmas1}) and (\ref{S-eq:modsigmas2}) for an electron-doped material. The parameter set is
    $n=0.85$, $t'=0.2t$ and a strain of magnitude $|\delta t/t| = 0.15$. (a)-(b) longitudinal response; (c)-(d) transverse
    response. All figures share a legend.} 
    \label{S-fig:CasesElecDoped} 
\end{figure}

In S-Fig.~\ref{S-fig:CasesHoleDoped} we gauge the sensitivity of different terms in Eqs.~(\ref{S-eq:modsigmas1}) and
(\ref{S-eq:modsigmas2}) to the effects of strain on the normalized longitudinal and transverse resistivity response as a
function of temperature for a hole-doped material at optimal density for both a compressive and tensile strain.
Additionally we show for comparison the resistivity calculation on the unstrained tetragonal lattice denoted ``Tetra''
and we show a calculation of the resistivity response on the strain-induced orthorhombic lattice called ``Ortho''.  The
latter curve combines the effects of strain from each term. We note that the strain-induced effects from the change in
the lattice constant is small compared to the other terms. We obverse that the longitudinal resistivity strain response
is dominated by the vertex contribution at high temperature, while the effect of $\Upsilon_{\vec{k}}$ becomes more and
more important in modifying the T-dependent behavior of the longitudinal response when decreasing the temperature, as
seen in panels (a) and (b).  The transverse resistivity, shown in panels (c) and (d), is mostly sensitive to the effect
of strain on the spectral functions, i.e., $\Upsilon_{\vec{k}}$ and therefore the interactions are the dominate source
of the response over the entire temperature range.

Now switching gears we examine the strain-induced effects on materials with electron doping, as shown in
S-Fig.~\ref{S-fig:CasesElecDoped} with parameters set to density $n=0.85$, second neighbor hopping $t'=0.2t$, and a
strain of magnitude $|\delta t/t| = 0.15$. In the electron-doped case we likewise find that the vertex term enhances the
longitudinal resistivity for compressive strain as shown in panel (a) whereas a tensile strain has the opposite effect. On
the other hand, we now find that the strain-induced effects contained in $\Upsilon_{\vec{k}}$ term enhances the resistivity at low-$T$
rather than suppressing the resistivity as seen in the hole-doped case. Continuing with the similarities we find that the
effects of strain on transverse response from vertex part remains negligible as seen in panels (c)-(d) and therefore the
strain-induced effects due to interactions dominate the resistivity calculation. Moreover, we observe a switch in the
sign of the response as we move from hole-doped to electron-doped which follows from the dominance of the strain-induced
effects on the correlations.

\begin{figure}[h]
    \includegraphics[width=.99\columnwidth]{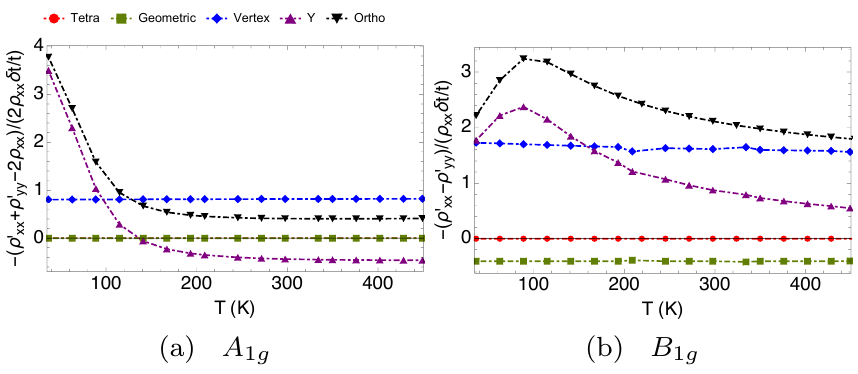}
    \caption{The strain-induced resistivity response vs $T$ for (a) $A_{1g}$ and (b) $B_{1g}$ irreps, isolating different
        terms in Eqs.~(\ref{S-eq:modsigmas1}) and (\ref{S-eq:modsigmas2}), for a material with hole doping. The parameter
        set is $n=0.85$, $t'=-0.2t$, with a compressive strain $\delta t/t = -0.15$. All figures share a legend.} 
    \label{S-fig:CasesA1gB1g} 
\end{figure}

In Figure~\ref{S-fig:CasesA1gB1g} we have the response function for the irrep $A_{1g}$ and
$B_{1g}$, respectively \cite{S-Fisher2015}. As one might expect, these curves are sensitive to both the vertex part and
the spectral function. We also note that the source of the $T$-dependence is the spectral function and that the dominate
source of asymmetry between compressive and tensile strains for the $A_{1g}$ and
$B_{1g}$ irreps is the vertex part at high-$T$ and the $\Upsilon_{\vec{k}}$ term at low-$T$.

\section{Kinetic Energy}
\label{S-sec:KineticEnergy}

%Now we look at total kinetic energy for an ECFL under a strain along the x-axis. Recall that the anisotropic kinetic energy can be related to a measurement of the optical weight through the f-sum rule\cite{S-optical1,S-optical2} on the \tJ model, and it is an experimentally accessible observable.

%The total kinetic energy for a system under strain is computed as
%\begin{equation} \label{S-eq:totkinetic}
%    K_{\text{tot}} = \Big \langle \int^{\infty}_{-\infty}\rho_\G(k) \epsilon_{\vec{k}} d\omega \Big \rangle_k\;,
%\end{equation}
%which may be decomposed as follows:
%\begin{equation}
%    K_{\text{tot}} = K_{xx} + K_{yy} + K_{xy},
%\end{equation}
%where $K_{xy}$ is the kinetic energy along the diagonal of orthorhombic lattice and is isotropic for each diagonal. 

%\begin{itemize}
%\item We present $A_{1g}$: $$\frac{K'_{xx} + K'_{yy} - 2K_{xx}}{2K_{xx} \delta t / t} \; \text{vs} \; T $$  
%\item We present $B_{1g}$ : $(K'_{xx} - K'_{yy}) / (K_{xx} \delta t / t) \; \text{vs} \; T$
%\item We present $XX$ : $(K'_{xx} - K_{xx}) / (K_{xx} \delta t / t) \; \text{vs} \; T$
%\item We present $YY$ : $(K'_{yy} - K_{yy}) / (K_{xx} \delta t / t) \; \text{vs} \; T$
%\end{itemize}

\subsection{The raw kinetic energies}

\begin{figure}[h]
    \includegraphics[width=.99\columnwidth]{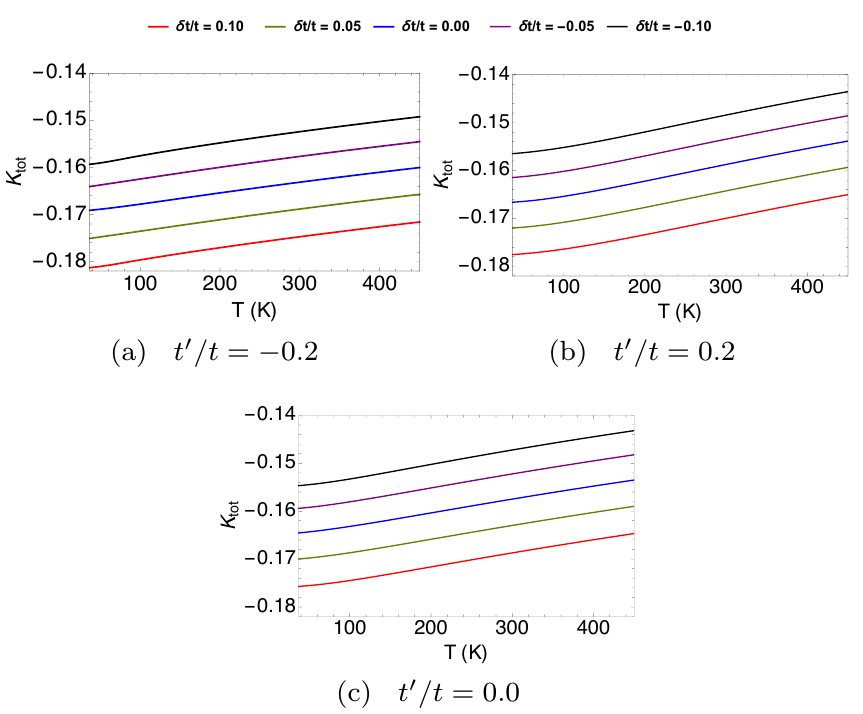}
    \caption{The total kinetic energy at optimal filling $n=0.85$, at various $t'$, for various $\delta t/t$. All
    figures shared a legend.} 
    \label{S-fig:totalkinetic} 
\end{figure}

\figdisp{S-fig:totalkinetic} shows the total kinetic energy $K_{\tx{tot}}$ versus $T$ at various $t'$ and $\delta t/t$.
The total kinetic energy for a system under strain is computed as
\begin{equation} \label{S-eq:totkinetic} 
    K_{\text{tot}} = \Big \langle \int^{\infty}_{-\infty}\rho_\G(k) \epsilon_{\vec{k}} d\omega \Big \rangle_k\;. 
\end{equation}
We observe that decreasing the strain at a fixed $T$ increases the total kinetic energy for all $t'$ which indicates
that the scale of the optical conductivity increases as the strain gets weaker. Recall that in the \tJ model the
kinetic energy is related to optical conductivity through the f-sum rule \cite{S-optical1,S-optical2} \big (see Eq.~(53) of the
main text\big). We also note that the curvature of the line shape, similar to $\rho$ - $T$ curves, slightly changes from negative to
positive as $t'$ increases from negative (hole-like) to positive (electron-like).

\begin{figure}[ht]
    \includegraphics[width=.99\columnwidth]{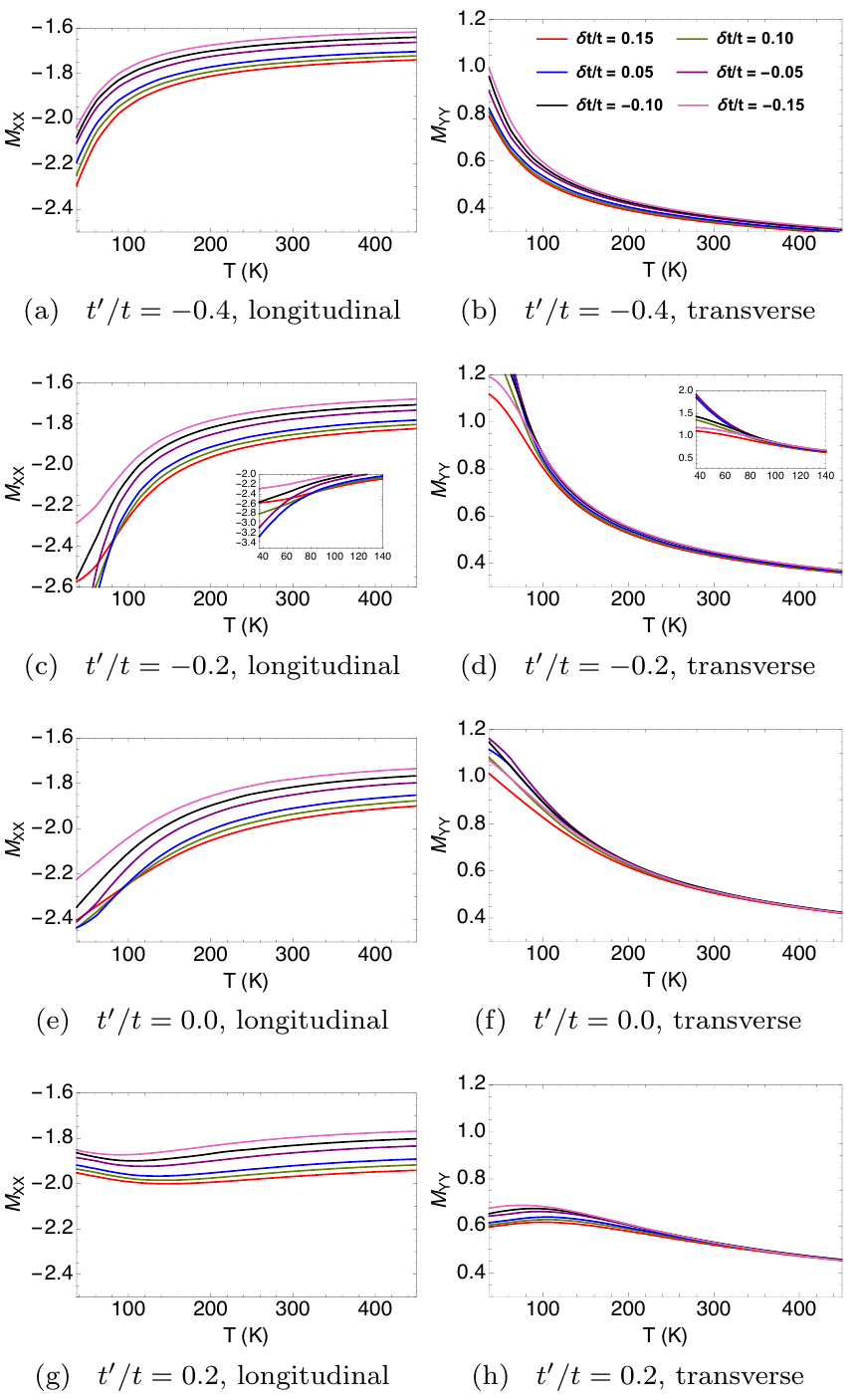}
    \caption{The normalized strain-kinetic-energy susceptibilities for longitudinal and transverse
     components from \disp{S-long-MXX} and \disp{S-trans-MYY}, respectively,  at optimal doping $n=0.85$, for various
    $t'$ and $\delta t/t$. All figures share a legend.}
    \label{S-fig:susKxxKyynd085}
\end{figure}

\subsection{The longitudinal and transverse susceptibilities}

The strain-kinetic-energy susceptibility for the longitudinal and transverse directions are defined in terms of hopping
strain as
\begin{eqnarray}
    M_{XX} \equiv -\Big ( \frac{K'_{xx}-K_{xx}}{K_{xx}} \Big ) 
        \Big / \Big (\frac{\delta t}{t} \Big ) \;, \label{S-long-MXX} \\
    M_{YY} \equiv -\Big (\frac{K'_{yy}-K_{yy}}{K_{xx}} \Big ) 
        \Big / \Big (\frac{\delta t}{t} \Big ) \;, \label{S-trans-MYY}
\end{eqnarray}
respectively. We plot the longitudinal and transverse strain-kinetic-energy susceptibilities verses temperature as shown
in S-Fig.~\ref{S-fig:susKxxKyynd085} for the optimal doping ($n=0.85$) case at various $t'$ and $\delta t / t$. 
%First, we note that increasing the density towards the Mott-insulating limit leaves the line shape of the strain curves intact, however, the strength of tends to increase, in particular that of the transverse response at low-$T$. 
Similar to the results for the resistivity response functions we find that the transverse response functions are nearly
linear at high-$T$ whereas the longitudinal response functions are non-linear throughout the temperature range and as
expected the intensity of the transverse response function is weaker in comparison to the longitudinal.  
%\sout{We also note that there is an approximate mirror symmetry between the longitudinal response and the transverse response. Although, the intensity of the transverse response is weaker which is consistent with the raw data. Now, let us focus on the longitudinal response at optimal density. We see that for all $t'$ the longitudinal response is non-linear at all displayed temperatures. Next, we examine the transverse response function which is linear at room temperature and becomes increasingly non-linear as the system is cooled.}

Finally, looking at the strain dependence we find that the susceptibilities increase monotonically as strain is varied from
compressive to tensile for nearly all $t'$ with notable exception of $t'=-0.2t$ (approximately the parameter for LSCO)
which at low-$T$ the response function increases monotonically with the intensity of the strain with a slight asymmetry
between tensile and compressive strains. For all other $t'$ we see, e.g. in \figdisp{S-fig:susKxxKyynd085} (f), that the
transverse kinetic energy cannot resolve between different tensile strains ($\delta t/t<0.0$) until very low
temperatures whereas it is quite distinguishable for the longitudinal case and likewise for the compressive strains.

\begin{figure*}[ht]
    \includegraphics[width=\textwidth]{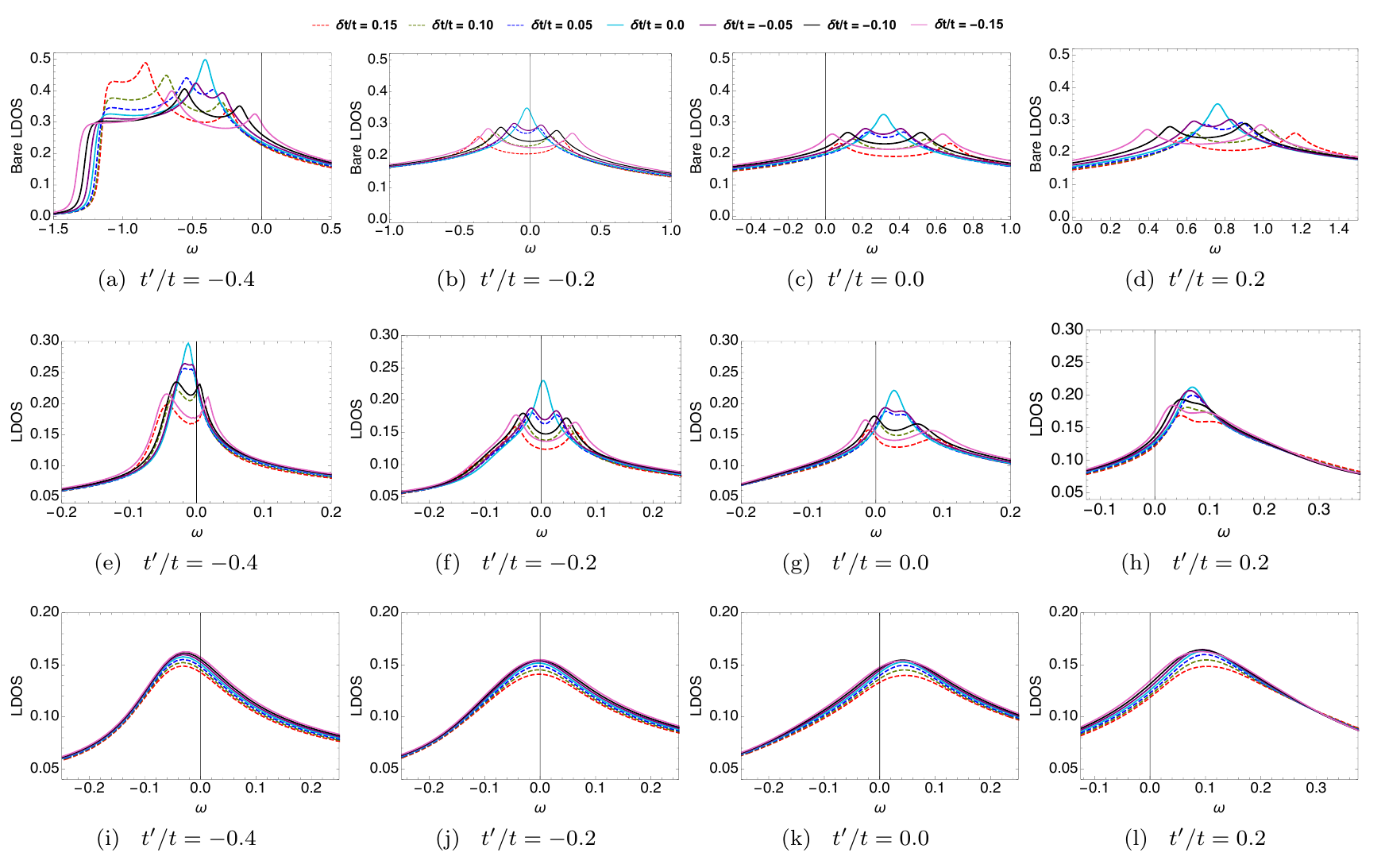}
    \caption{The local density of states versus frequency at optimal filling $n=0.85$, at $T=37$K and room temperature
        ($T=298$K), for various $t'=-0.4,-0.2,0.0,0.2$ at select $\delta t/t$. (a)-(d) The non-interacting system (i.e., band structure), $T=298$K;
    (e)-(h) the interacting system, $T=37$K; (i)-(l) the interacting system, $T=298$K.  All figures share a legend.}
    \label{S-fig:LDOSnd085}

\end{figure*}

\section{The Local Density of States : Strain Variation}
\label{S-sec:LDOS}

In S-Figs.~\ref{S-fig:LDOSnd085}, we present the LDOS at optimal density ($n=0.85$) at $T=37$K and room temperature
($T=298$K) for various $t'$ and $\delta t / t$ for systems with and without interactions.
%\sout{Comparing these two figures, we note that the density has little effect beyond reducing the height of interacting system as it increases towards the Mott-insulating limit. Therefore, we will focus on analyzing \figdisp{S-fig:LDOSnd080} since it applies to \figdisp{S-fig:LDOSnd085} as well.} 
In panels (a)-(d) we have the bare LDOS where we find that strain causes a highly localized inversion of the peak,
resulting in a valley centered at same approximate position along the spectrum. The valley is surrounded by a peak on
each side at a reduced height and the weight of spectrum is lifted to the background. 

For $t'/t < 0$ ($t'/t > 0$) the bare LDOS peak heights are asymmetric, higher on the left (right), and shifted left
(right) on the spectrum, whereas for $t'/t = 0$ the LDOS peaks are symmetric. We also notice an asymmetry between a
compressive strain ($\delta t / t > 0$) and a tensile strain ($\delta t / t < 0$) where the bare LDOS peak heights are
more strongly inverted for the former except for $t'/t = -0.4$ since it is constrained by the band edge. 

We see that at low temperatures, i.e. $T=37$K as seen in panels (e)-(h), the double peak persists for all strains,
however, the peaks remain asymmetric favoring the left hand side for all $t'$. In previous work \cite{S-PS}, it was
shown that systems with high $t'$ have higher Fermi liquid temperature scales, making them more robust to warming, that
is, the LDOS peaks will have a slower rate of change (smoothening and broadening) in response to warming. Although we
may expect the Fermi liquid temperature scale to warm as the hopping strain increases from negative (tensile) to
positive (compressive), we find that the scale is similar for compressive and tensile strains of identical magnitude as
seen in panels (i)-(l). This is a result of the anisotropy of the resistivity, producing two channels and hence two
Fermi temperature scales. We find that the rate of change in LDOS peaks scales with the average ratio of the hopping
parameters $(t_d / t_x + t_d / t_y) / 2$, where $t_d / t_x$ and $t_d / t_y$ correspond with the Fermi temperature scale
for the longitudinal and transverse resistivity, respectively.

\end{document}